\documentclass[twocolumn]{aastex63}

\usepackage[T1]{fontenc}
\usepackage{aas_macros}
\usepackage{natbib}
\usepackage{times}

\newcommand{\package}[1]{\textsl{#1}}

\newcommand{\msun}{\mbox{M$_\odot$}}

\newcommand{\pc}{\mbox{${\rm pc}$}}

\newcommand{\kpc}{\mbox{${\rm kpc}$}}
\newcommand{\kms}{\mbox{${\rm km}~{\rm s}^{-1}$}}

\newcommand{\feh}{\mbox{$[{\rm Fe}/{\rm H}]$}}
\newcommand{\mgfe}{\mbox{$[{\rm Mg}/{\rm Fe}]$}}

\newcommand{\vphi}{\mbox{$v_\phi$}}
\newcommand{\vz}{\mbox{$v_{\rm z}$}}
\newcommand{\mpl}{\mbox{$m_{\rm p}$}}

\newcommand{\me}{\mbox{M$_\oplus$}}

\newcommand{\hjcj}{\mbox{${\rm HJ}:{\rm CJ}$}}

\newcommand{\be}{\begin{equation}}
\newcommand{\ee}{\end{equation}}
\newcommand{\bea}{\begin{eqnarray}}
\newcommand{\eea}{\end{eqnarray}}

\received{2021 September 6}
\revised{}
\accepted{}

\submitjournal{ApJ}
\shorttitle{Galactic Dynamics Shape Planetary Systems}
\shortauthors{Kruijssen et al.}

\usepackage{amsmath}

\begin{document}\sloppy\sloppypar\raggedbottom\frenchspacing

\title{\vspace{-4mm}Not the Birth Cluster: the Stellar Clustering that Shapes Planetary Systems\\ is Generated by Galactic-Dynamical Perturbations}

\correspondingauthor{J.~M.~Diederik~Kruijssen}
\email{kruijssen@uni-heidelberg.de}

\author[0000-0002-8804-0212]{J.~M.~Diederik~Kruijssen}
\affil{Astronomisches Rechen-Institut, Zentrum f\" ur Astronomie der Universit\"at Heidelberg, M\"onchhofstra\ss e 12-14, D-69120 Heidelberg, Germany}

\author[0000-0001-6353-0170]{Steven~N.~Longmore}
\affil{Astrophysics Research Institute, Liverpool John Moores University, IC2, Liverpool Science Park, 146 Brownlow Hill, Liverpool L3 5RF, UK}

\author[0000-0002-5635-5180]{M\'{e}lanie~Chevance}
\affil{Astronomisches Rechen-Institut, Zentrum f\" ur Astronomie der Universit\"at Heidelberg, M\"onchhofstra\ss e 12-14, D-69120 Heidelberg, Germany}

\author[0000-0003-3922-7336]{Chervin~F.~P.~Laporte}
\affil{Institut de Ci\`{e}ncies del Cosmos (ICCUB), Universitat de Barcelona (IEEC-UB), Mart\'{i} i Franqu\`{e}s 1, 08028 Barcelona, Spain}

\author[0000-0002-4764-7130]{Michal Motylinski}
\affil{Astrophysics Research Institute, Liverpool John Moores University, IC2, Liverpool Science Park, 146 Brownlow Hill, Liverpool L3 5RF, UK}

\author[0000-0002-9642-7193]{Benjamin~W.~Keller}
\affil{Astronomisches Rechen-Institut, Zentrum f\" ur Astronomie der Universit\"at Heidelberg, M\"onchhofstra\ss e 12-14, D-69120 Heidelberg, Germany}

\author[0000-0001-9656-7682]{Jonathan~D.~Henshaw}
\affil{Max-Planck Institut f\"{u}r Astronomie, K\"{o}nigstuhl 17, 69117 Heidelberg, Germany}


\begin{abstract}\noindent
Recent work has demonstrated that exoplanetary system properties correlate strongly with ambient stellar clustering in six-dimensional stellar position-velocity phase space, quantified by dividing planetary systems into sub-samples with high or low phase space densities (`overdensity' and `field' systems, respectively). We investigate the physical origins of the phase space overdensities and, thereby, which environmental mechanisms may have impacted the planetary systems. We consider the galactic-scale kinematic structure of the Milky Way observed with \textit{Gaia} and show that the overdensities correspond to the well-known, kpc-scale kinematic ripples and streams in the Galactic disk, which are thought to be generated by bar and spiral arm-driven resonances and satellite galaxy passages. We also find indications that the planet demographics may vary between individual phase space overdensities, which potentially have differing physical origins and histories. Planetary systems associated with the `phase space spiral' (a recent perturbation of the Galactic disk) have a hot Jupiter-to-cold Jupiter ratio that is 10 times higher than in field systems. Finally, the hot Jupiter-to-cold Jupiter ratio within overdensities may increase with host stellar age over Gyr timescales. Because the overdensities persist for several Gyr, we argue that late-time perturbations of planetary systems most likely explain these trends, although additional perturbations at birth may contribute too. This suggests that planetary system properties are not just affected by stellar clustering in their immediate surroundings, but by galaxy-scale processes throughout their evolution. We conclude by discussing the main open questions towards understanding the diversity of physical processes that together set planetary system architectures.
\end{abstract}

\section{Introduction}
\label{sec:intro}
\subsection{Environmental effects on planetary systems}
It is a long-standing question whether most planetary systems exist in isolation, such that their formation and evolution can be described by physical processes internal to the system \citep[e.g.][]{benz14}, or that they are affected by the ambient stellar and galactic environment \citep[e.g.][]{adams10}. Since the discovery of the external photoevaporation of protoplanetary disks in the Orion Nebula Cluster (`proplyds') by \citet{odell94}, it has become clear that external irradiation may affect the planet formation process in at least some environments, as has since been corroborated by two decades of modelling \citep[e.g.][]{johnstone98,armitage00,scally01,adams04,clarke07,winter20} and observations \citep[e.g.][]{dejuanovelar12,ansdell17,vanterwisga19}. Similarly, there has been a long history of models quantifying the dynamical perturbation of protoplanetary disks and planetary systems by stellar encounters in their birth cluster \citep[e.g.][]{clarke93,olczak06,malmberg11,rosotti14,cai18,fujii19}, by encounters with field stars \citep[e.g.][]{zakamska04}, or by the galactic large-scale tidal field \citep[e.g.][]{kaib13}.

Despite these encouraging demonstrations that the large-scale stellar environment can indeed affect planetary systems, the fraction of systems affected by such perturbations long remained unclear. Using the revolutionary astrometry provided by ESA's \textit{Gaia} satellite \citep{gaia16,gaia18}, a series of recent papers has shown that a significant fraction of planetary systems have likely been shaped by their ambient stellar environment. By using \textit{Gaia} to measure the ambient stellar phase space density of exoplanetary systems and dividing the planet sample into environments of low and high ambient stellar phase space density (which we refer to as planets residing in the `field' and in `overdensities', respectively), \citet{wklc20} found that planetary systems in overdensities exhibit significantly shorter orbital periods and enhanced hot Jupiter occurrence rates. Subsequently, it was found that overdensity systems have decreased planetary multiplicities \citep[implying that environmental effects may be partially responsible for the ``Kepler dichotomy'', \citealt{lissauer11}]{longmore21}, an elevated ratio of super-Earths to sub-Neptunes \citep[changing their distribution around the ``radius valley'', \citealt{fulton17}]{kruijssen20d}, and an increased degree of planetary radius uniformity within single systems \citep[i.e.\ ``peas in a pod'', \citealt{weiss18}]{chevance21}. These results have been confirmed and expanded by independent studies from other groups, focusing primarily on planetary multiplicity and eccentricity \citep{dai21} and giant planet occurrence and stellar age \citep[also see Section~\ref{sec:disc} for a brief discussion on the possible covariance of planet properties with stellar age and chemistry]{adibekyan21,mustill21}.

These results empirically demonstrate that the current degree of stellar clustering in position-velocity phase space affects the architectures and properties of a significant (and possibly dominant) fraction of known exoplanetary systems. This implies that the formation and evolution of a planetary system cannot be fully described by only considering processes internal to the planetary system. However, it is unclear which physical mechanisms are responsible. The answer to this question requires knowledge of the physical origins of the phase space overdensities. Establishing what types of structures these represent would enable assessing when and through which environmental mechanisms the planetary systems might have been affected.

\subsection{Physical origins of the phase space overdensities}
The key question is what are the physical origins of the phase space overdensities -- its answer would shed light on the environmental mechanisms that may have impacted the associated planetary systems. If the overdensities represent dispersed birth clusters, then the associated planetary systems may have experienced external photoevaporation from nearby massive stars. However, if the overdensities are generated by galactic-dynamical processes, then late-time perturbations (e.g.\ by stellar encounters or galactic tides) may explain the observed environmental dependence.

The phase space overdensities were identified by \citet{wklc20} by considering all stars with full 6D position-velocity phase space information from \textit{Gaia} that reside within 40~pc of an exoplanetary system. For a randomly-drawn subset of 600 stars in this volume, the normalized phase space density was obtained by measuring the ‘Mahalanobis distance’ to the 20${\rm th}$-nearest neighbour in phase space, inverting the resulting 6D hypervolume spanned by that distance, and dividing it by the median of all drawn stars. The quantity $P_{\rm null}$ was then defined to represent the probability that the resulting distribution of phase space densities of the stars within 40~pc of the planetary system is described by a unimodal Gaussian distribution. Most of the phase space density distributions are not well described by a single log-normal ($P_{\rm null}<0.05$), and for each of these systems \citet{wklc20} computed the probability that the star is associated with the low- or high-phase space density component ($P_{\rm low}$ and $P_{\rm high}\equiv1-P_{\rm low}$, respectively) using Gaussian mixture modelling. When considering the distributions of stars with high-confidence associations (`field' with $P_{\rm low}>0.84$ and `overdensity' with $P_{\rm high}>0.84$) in physical ($\{x,y,z\}$) space over the 40-pc radius region, both populations are fully mixed and no substructure can be discerned. However, when considering the field and overdensity populations in velocity ($\{v_x,v_y,v_z\}$) space, both components are clearly segregated, implying that the phase space overdensities represent co-moving groups within the galactic disk \citep[see e.g.][]{eggen65,dehnen98,famaey05}.

A tempting interpretation may be that the phase space overdensities are the remnants of the birth clusters within which the planetary systems formed, and the differences in planetary demographics relative to field systems reflect the impact of the birth cluster \citep[e.g.][]{rodet21}. After all, there is a wealth of evidence that the external photoevaporation of protoplanetary disks takes place in nature, and the stellar encounter rate is orders of magnitude greater within birth clusters than in the field \citep[e.g.][]{zakamska04}. However, such an enhancement of the stellar encounter rate would require gravitational boundedness, because unbound stellar associations only live for a crossing time \citep{gieles11}. In Milky Way-like galaxy disks, only 5--10\% of the star formation occurs in gravitationally-bound clusters \citep[e.g.][]{goddard10,kruijssen12d,adamo20}, which implies that stellar encounters in the birth cluster cannot explain the observed dependence of planet properties on stellar clustering in phase space. Irrespectively of whether the birth cluster is gravitationally bound or unbound, it will disperse into the field of the galactic disk on timescales of $10^7{-}10^9$~yr \citep[e.g.][]{krumholz19}. This decreases the likelihood that the close ($<40$~pc) environments of the 1-4.5~Gyr old planetary systems analysed by \citet{wklc20} are still dominated by the co-moving remains of the clusters and associations within which the planetary systems may have formed.

Alternatively, it is possible that the phase space overdensities have been generated by processes unrelated to the natal stellar clustering, but induced by galactic dynamics. The phase space distribution of stars in the Galactic disk revealed by \textit{Gaia} is highly substructured, exhibiting waves, ripples, and streams \citep[e.g.][also see e.g.\ \citealt{widrow12} for the pre-\textit{Gaia} perspective]{antoja18,kawata18,ramos18,schoenrich18,alves20}. These structures are thought to be generated by resonances and instabilities driven by the bar \citep[e.g.][]{fragkoudi19,laporte20} and spiral arms \citep{hunt18,quillen18}, or by the passages of satellite galaxies through the galactic disk \citep[e.g.][]{laporte19,hunt21}, which can simultaneously drive spiral and vertical perturbations \citep[e.g.][]{edelsohn97,purcell11,gomez13,donghia16,laporte18}. Within these structures, the velocities of the stars are correlated over distances of several kpc. It is not implausible that the stellar encounter rate is enhanced in these co-moving groups, which might result in a correlation between planetary system properties and the stellar phase space density. If the overdensities identified by \citet{wklc20} do indeed represent these galactic-dynamical features, their large spatial extent would also naturally explain why the overdensities are not visible in physical ($\{x,y,z\}$) space within length scales of $\sim100$~pc and can only be detected in velocity space. However, even in overdensities the encounter rate with other stars is low and close encounters are rare, such that the timescale for sufficiently perturbative encounters may be longer than the ages of the planetary systems considered. An alternative option might be that planetary systems in overdensities experience enhanced external tidal perturbations (e.g.\ from nearby stars, gaseous structures, or the Galactic disk), which might shape them over Gyr timescales.

In this paper, we compare the kinematics of exoplanet host stars in overdensities and in the field to the kinematics of all stars with 6D phase space information in \textit{Gaia}'s second data release \citep[DR2,][]{gaia18} to determine whether the overdensities correspond to the galactic-dynamical features previously identified in the \textit{Gaia} data. We find close agreement with these features, which indicates that the overdensities that shape planetary system properties do not represent the remains of the birth cluster, but instead were generated by satellite galaxy passages or resonances driven by the bar or spiral arms. These findings imply that stellar clustering plays an intermediate, but central role in a much larger, multi-scale chain of causally-connected physical processes, in which the physics of galaxy formation and evolution drive variations in stellar clustering, which in turn shapes the properties of planetary systems \citep[also see][]{kruijssen20}.

\section{Observational data}
\label{sec:data}
\subsection{Planetary system sample}
\citet{wklc20} calculated the relative position-velocity phase space densities for all known exoplanet host stars in the \citet{exoarchive} that have radial velocities from \textit{Gaia} DR2 \citep{gaia18}. At the time of sample construction (May 2020), this restricted the sample to 1525 out of 4141 confirmed exoplanets. Additionally, exoplanet host stars with less than 400 neighbours with 6D phase space information within 40~pc, $P_{\rm null}\geq0.05$, or $0.16\leq P_{\rm low/high}\leq0.84$ are considered to have an ambiguous phase space density classification and are removed from the sample, leaving a total of 716 planetary systems.

The sample of planetary systems with a phase space density classification in \citet{wklc20} is restricted further using the following sample cuts. Systems with ages younger than 1~Gyr are omitted, to exclude planetary systems that might not yet have stabilized after their formation \citep[e.g.][]{kennedy13}. Systems older than 4.5~Gyr are excluded, because empirically the occurrence rate of overdensities drops precipitously at older ages. Additionally, the host stellar masses are restricted to a narrow interval of $M_{\rm s}=0.7{-}2.0~\msun$ to avoid indirectly probing any dependence on the host mass. This leaves a final sample of 284 planetary systems, with 45 systems residing in the field and 239 residing in overdensities, containing 60 and 308 known planets, respectively.

\begin{figure*}
\includegraphics[width=\hsize]{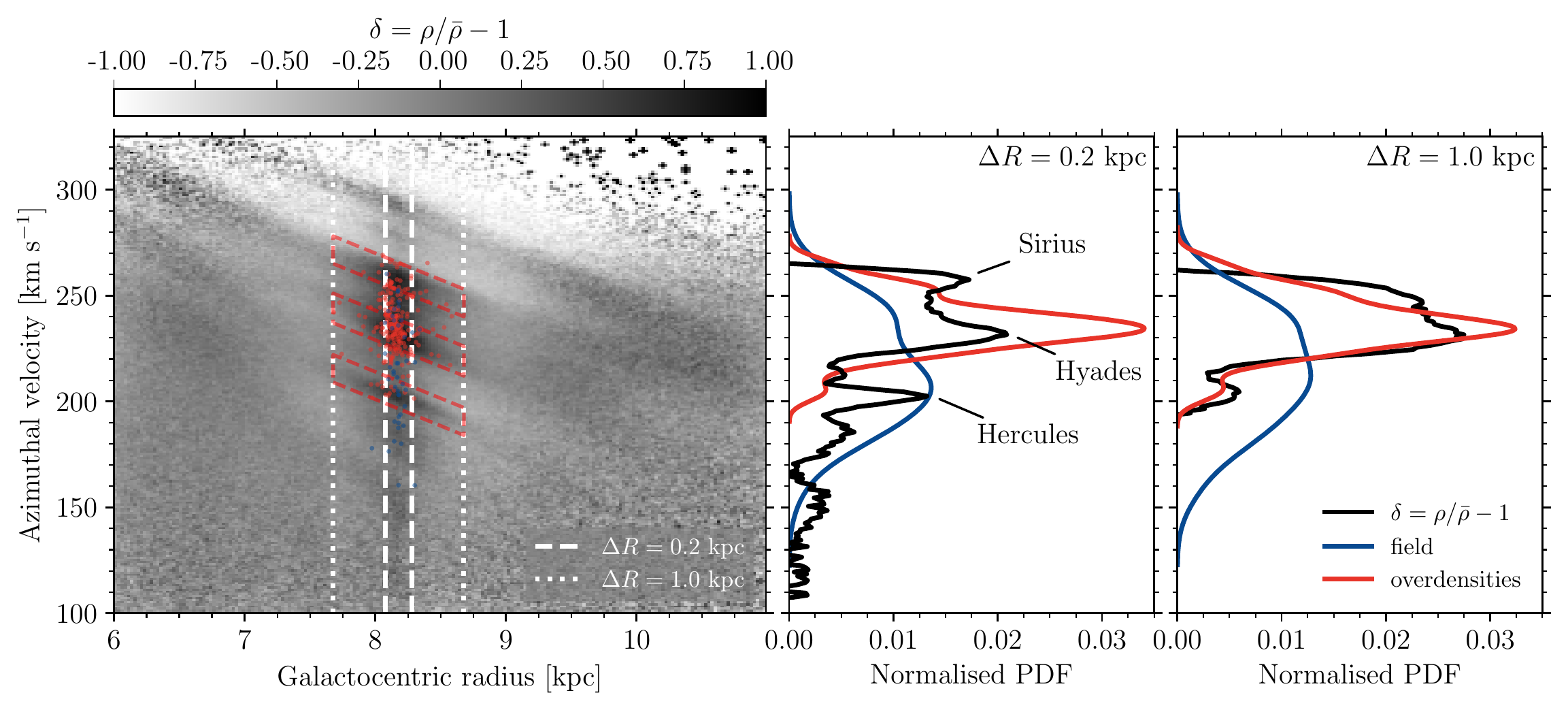}%
\caption{
\label{fig:rvphi}
Distribution of stellar azimuthal velocities as a function of galactocentric radius (left). The gray scale shows a two-dimensional, unsharp-masked histogram of all stars for which radial velocities (and thus the full 6D phase space information) is available from \textit{Gaia}. The unsharp masking is performed to highlight the local phase space density contrast. The distribution of \textit{Gaia} stars clearly illustrates the ridges that span several kpc and dominate the local phase-space structure of the Galactic disk. The positions of the exoplanet host stars are indicated by the \{blue, red\} symbols for \{field, overdensity\} systems. The systems in overdensities are strongly correlated with the kpc-scale dynamical features in gray. Red dashed lines indicate the selection criteria for planetary systems associated with individual features (see \autoref{fig:planets_streams}). The vertical white lines indicate the two galactocentric radius intervals for which the middle and right-hand panels show the normalized PDFs of the azimuthal velocity for the full \textit{Gaia} sample, together with the exoplanetary systems in phase space overdensities and in the field. Also in this projection, the overdensity systems are clearly associated with the kpc-scale, galactic-dynamical features generated by bar, spiral arm, or satellite perturbations (several of which are indicated in the middle panel), whereas the field systems do not exhibit this association.
}
\end{figure*}

\subsection{Stellar position-velocity phase space}
To characterize the stellar kinematics in the Galactic disk, we use the full sample of stars with radial velocities from \textit{Gaia} DR2. We elect to use DR2 over EDR3 for consistency with the phase space classification of the planetary system sample from \citet{wklc20}. Stars with parallaxes smaller than 4.5 times the parallax uncertainty are removed \citep{rybizki21}. This yields a sample of just over 7 million stars. The phase space coordinates are transformed from the international celestial reference system (ICRS) to galactocentric coordinates by assuming a distance to the Galactic Center of 8178~pc \citep{gravity19} and a position of the Sun at 15~pc above the Galactic plane, a local circular velocity of $240~\kms$, and a Solar motion relative to the local standard of rest of $\{U_\odot,V_\odot,W_\odot\}=\{11.10,12.24,7.25\}~\kms$ \citep[following][]{schoenrich10}. Our results are unaffected by reasonable changes of these choices.

\section{Distribution of planetary systems in\break position-velocity phase space}
\label{sec:results}

\subsection{Association with kpc-scale dynamical structures}

\label{sec:streams}
In the left-hand panel of \autoref{fig:rvphi}, we show the two-dimensional distribution of the 7 million stars with 6D phase space information from \textit{Gaia} in the galactocentric radius-azimuthal velocity ($R{-}\vphi$) plane, together with the planetary systems, colored by their phase space density classification. The stellar sample is strongly biased towards radii close to the solar circle, which complicates the identification of phase space structure. In \autoref{fig:rvphi}, we divide out the selection function and emphasize the local contrast by applying an unsharp mask to the two-dimensional histogram \citep[following][]{laporte19}. We bin the data using bins with size $\{\delta R,\delta \vphi\}=\{25~\pc,1~\kms\}$ and generate the unsharp mask by dividing high- ($\rho$) and low-resolution ($\bar{\rho}$) renditions of the histogram. These are obtained by convolving with a Gaussian kernel using a standard deviation of 0.5 and 12 pixels, respectively. The quantity shown in \autoref{fig:rvphi} is the normalized ratio of both histograms $\delta=\rho/\bar{\rho}-1$, such that $\delta=0$ corresponds to $\rho=\bar{\rho}$.

The long, kpc-scale kinematic features in \autoref{fig:rvphi} immediately catch the eye. These ridges are well known \citep[e.g.][]{antoja18,quillen18,fragkoudi19,laporte19,laporte20} and represent co-moving groups that are triggered by the time-dependent nature of the gravitational potential. The precise mechanism is unclear, but the features have been reproduced in numerical models as products of resonances and instabilities caused by the bar, spiral arms, and satellite galaxy passages. These seem to be long-lived features that can persist over several Gyr and can be rejuvenated following satellite galaxy impacts. The features dominate the phase space structure of the Galactic disk, as is illustrated by the fact that $\delta$ has absolute values of order unity.

\begin{figure*}
\centering
\includegraphics[width=\hsize]{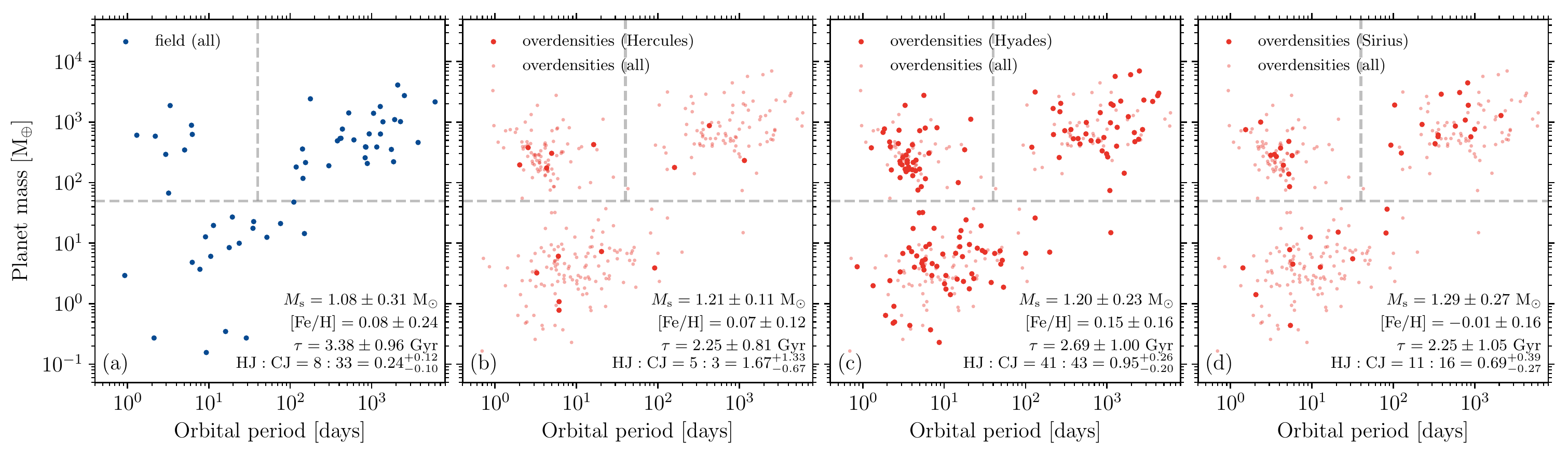}%
\caption{
\label{fig:planets_streams}
Masses and orbital periods of planets in field (panel~a) and overdensity (panels~b-d) systems, for all planets (small symbols) and those associated with individual galactic-dynamical features (large symbols; see the red dashed boxes in \autoref{fig:rvphi}). Gray dashed lines indicate the cuts used to select giant planets ($\mpl\geq50~\me$) and distinguish between hot ($P<40$~days) and cold Jupiters ($P\geq40$~days). The panels illustrate the characteristic difference between field and overdensity planets that we discovered in \citet{wklc20} and which motivated this study, but we now additionally find statistically significant differences in the hot Jupiter-to-cold Jupiter ratio (\hjcj; see annotations) between the sub-samples of planetary systems associated with individual dynamical features. The median and standard deviations of the host stellar masses $M_{\rm s}$, metallicities $\feh$, and ages $\tau$ are listed in the bottom-right corner of each panel and do not exhibit significant differences, showing that the variation of the \hjcj\ ratio does not result from covariance with host stellar properties (nor with distance, see \autoref{fig:distance_streams}).
}
\end{figure*}
The structures in the left-hand panel of \autoref{fig:rvphi} are best defined along the azimuthal velocity axis, because they extend over many kpc in radius. To visualize whether the planetary systems in phase space overdensities are correlated with these kpc-scale features, the middle and right-hand panels show the normalized azimuthal velocity distributions in two different galactocentric radius slices, of total widths 0.2~kpc and 1.0~kpc around the solar radius, respectively. These radius cuts are applied to avoid dilution by the $R{-}\vphi$ gradient visible in the left-hand panel. The resulting one-dimensional histograms of the stellar distribution are constructed by summing $\delta$ (including possible negative values) over the spanned range of galactocentric radii and subsequently setting any $\vphi$ bins with negative values to zero before normalising. This means that the black lines are effectively showing the normalized PDF only of the phase space overdensities in the stellar sample. For comparison, we apply a kernel density estimate to the distributions of the planetary systems to minimize any dependence on binning due to the low-number statistics involved in the tails of the planetary system distributions.

Taken together, the panels in \autoref{fig:rvphi} clearly show that the planetary systems in overdensities are correlated with the kpc-scale ridges seen in the stellar sample. Using the galactocentric radius slice of $\Delta R=1.0$~kpc (right-hand panel), we see that the azimuthal velocity distributions of overdensity systems and the phase space overdensities of the full stellar sample are nearly indistinguishable. For a radius slice of $\Delta R=0.2$~kpc (middle panel), the contrast is enhanced and the individual features are more visible. We see that most of the overdensity systems are associated with the Hyades moving group, but the red histogram has shoulders at the azimuthal velocities of the Sirius and Hercules moving groups, indicating that a subset of the overdensity systems is associated with these features.

Thus far, it was unknown whether the stellar clustering in phase space around planetary systems reflected a single co-moving structure within the Galaxy or a large number of features, each unique to the planetary system they are associated with. For this reason, studies to date have adopted a binary phase space classification into field and overdensity systems \citep{kruijssen20d,wklc20,adibekyan21,chevance21,dai21,longmore21,mustill21}. \autoref{fig:rvphi} shows that the true answer likely lies somewhere in between -- there are several co-moving structures within the Galaxy, each of which is associated with multiple planetary systems. Now that we have identified at least three such different phase space structures hosting planetary systems, it is useful to consider the planet demographics for each of these individually.

In \autoref{fig:planets_streams}, we show the distribution of planets in the orbital period-planet mass ($P{-}\mpl$) plane, both for field systems and for systems associated with the three ridges identified above (Hercules, Hyades, and Sirius; defined by eye using the red dashed boxes in \autoref{fig:rvphi}). The different panels clearly illustrate that the occurrence rate of hot Jupiters varies strongly between samples; given that this is one of the main differences between planets in overdensities and in the field \citep{wklc20}, it has been the main point of focus of several of the studies immediately following on the discovery \citep[e.g.][]{dai21,mustill21,winter21}. However, \autoref{fig:planets_streams} now additionally shows that the hot Jupiter-to-cold Jupiter ratio (\hjcj) may vary between different overdensities. While in the field we have $(\hjcj)_{\rm field}\approx0.25$ and the overdensity population at large shows $(\hjcj)_{\rm overdensities}\approx1.0$, we also obtain marginally significant differences between the overdensities, with $(\hjcj)_{\rm Hercules}=1.67^{+1.33}_{-0.67}$, $(\hjcj)_{\rm Hyades}=0.95^{+0.26}_{-0.20}$ and $(\hjcj)_{\rm Sirius}=0.69^{+0.39}_{-0.27}$. The uncertainties on these \hjcj\ ratios are obtained by adopting a binomial distribution for the number of hot Jupiters ($N_{\rm HJ}$), for a total number of giant planets $N_{\rm HJ}+N_{\rm CJ}$ and a probability of drawing a hot Jupiter of $p_{\rm HJ}=N_{\rm HJ}/(N_{\rm HJ}+N_{\rm CJ})$. We then use the 16$^{\rm th}$ and 84$^{\rm th}$ percentiles of the resulting binomial distribution to find the corresponding values of $N_{\rm HJ}/N_{\rm CJ}$. Using the same binomial experiment, we obtain probabilities that the above ratios are consistent with the ratio observed in the field of $P_{\rm Hercules}=9.3\times10^{-3}$, $P_{\rm Hyades}=1.5\times10^{-9}$, and $P_{\rm Sirius}=9.1\times10^{-3}$, respectively.

Given the respective total numbers of giant planets between these sub-samples, the Sirius overdensity has thrice the hot Jupiter occurrence of the field, and the occurrence increases by another factor of $\sim2$ for the Hyades and Hercules overdensities. Comparison of the distributions of host stellar masses, metallicities, and ages does not reveal any significant differences between the sub-samples (for reference, the median and standard deviation of these quantities are provided in \autoref{fig:planets_streams}). While \citet{wklc20} already demonstrated that the difference in planet properties between the field and the entire population of overdensities does not result from covariance with any differences in host stellar properties, this now demonstrates that the same applies when dividing up the planet population further, by its association with the Hercules, Hyades, and Sirius overdensities. Finally, the differences are also unlikely to result from differences in detection method, because the fraction of planets discovered by transit measurements is very similar between the Hercules (0.50), Hyades (0.38), and Sirius (0.41) overdensities. Nonetheless, our results should be followed up by a thorough detectability assessment, also folding in differences in the non-detection rate between the different surveys and detection methods \citep[e.g.][]{mulders19}.

\begin{figure}
\centering
\includegraphics[width=\hsize]{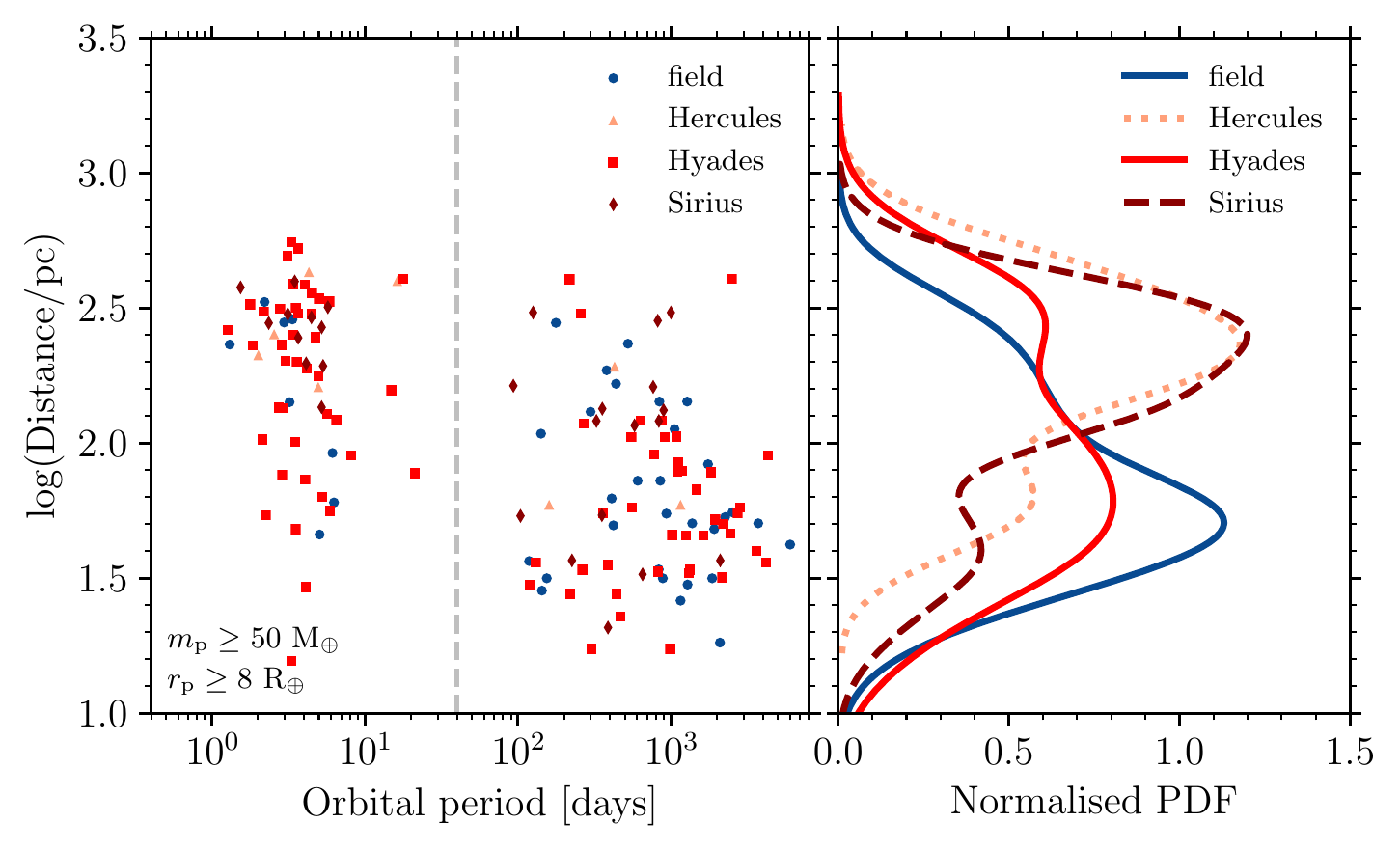}%
\caption{
\label{fig:distance_streams}
Distance distribution of giant planet host systems in the field and in the three main overdensities from \autoref{fig:rvphi}, as a function of orbital period (left) and as a kernel density estimate-smoothed histogram (right). The distance distributions of all sub-samples overlap, but their shapes differ. Importantly, the nearest overdensity (Hyades) has one of the highest hot Jupiter-to-cold Jupiter ratios (\hjcj) in \autoref{fig:planets_streams}. This difference is opposite to what would have been needed to explain the \hjcj\ variations between overdensities by a detection bias.
}
\end{figure}

The key remaining potential source of bias is the distance distribution of the planetary systems. Cold Jupiters are more challenging to detect due to their large semi-major axes and long orbital periods, and their detection thus requires smaller distances. \autoref{fig:distance_streams} shows the distribution of the giant planets in the orbital period-distance ($P{-}d$) plane for each of the four sub-samples in \autoref{fig:planets_streams}. The figure shows that the distance distributions do differ. On average, the field systems are closer than the overdensity systems, which indeed could potentially explain why \hjcj\ is higher in overdensities. However, as shown by \citet[see their Extended Data Figure~5]{wklc20}, the distance distributions of field and overdensity samples are statistically indistinguishable when only considering systems within $d\leq300$~pc, yet the difference in \hjcj\ between overdensities and the field persists. This means that a distance bias is not the source of the difference in hot Jupiter occurrence \citep[also see][]{dai21}. The remaining question is whether the difference in \hjcj\ between the individual overdensities, with $(\hjcj)_{\rm Hercules}>(\hjcj)_{\rm Hyades}>(\hjcj)_{\rm Sirius}$, may result from a distance bias. This would require the planetary systems associated with the Hercules overdensity to reside at the largest distances, followed by those in the Hyades overdensity, with planetary systems in the Sirius overdensity being situated closest to the Sun. However, as shown by \autoref{fig:distance_streams}, the Hyades systems are actually somewhat closer on average than the Sirius systems, while the Sirius and Hercules systems have nearly indistinguishable distance distributions. This means that the differences in \hjcj\ between the individual overdensities do not result from a distance bias.

Summarising the above discussion, we find that the stellar position-velocity phase space overdensities that have recently been found to correlate with the architectures of associated planetary systems correspond to kpc-scale ripples in the Galactic disk that are generated by galactic-dynamical processes such as resonances and instabilities. Moreover, we find that there exist detectable physical differences in planetary system architectures between different phase space overdensities. Given that the formation mechanisms and ages of these overdensities may differ too, this provides a promising and potentially important avenue to further constrain how planetary systems are affected by the galactic environment. We discuss this point further in \S\ref{sec:disc}.

\subsection{An overdensity induced by a satellite galaxy?}
\label{sec:spiral}
\begin{figure*}
\centering
\includegraphics[width=0.9\hsize]{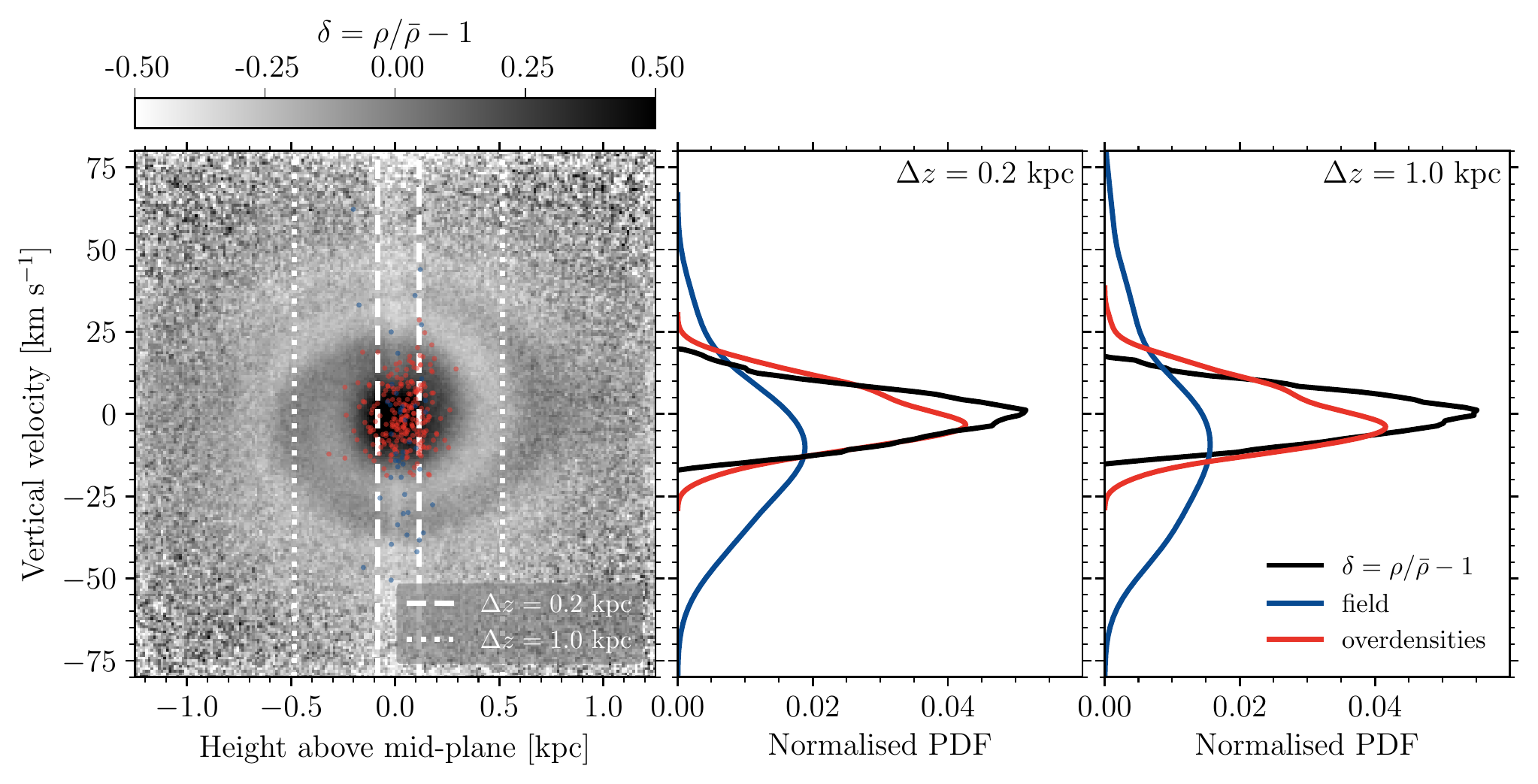}%
\caption{
\label{fig:zvz}
Distribution of stellar vertical velocities as a function of their height above the galactic mid-plane (left). As in \autoref{fig:rvphi}, the gray scale shows a two-dimensional, unsharp-masked histogram of all stars with 6D phase space information from \textit{Gaia}, which visualizes the phase space spiral that is thought to have been generated by the passage of the Sagittarius dwarf galaxy \citep{antoja18,laporte19}. The positions of the exoplanet host stars are indicated by the \{blue, red\} symbols for \{field, overdensity\} systems. The vertical white lines indicate the two height intervals for which the middle and right-hand panels show the normalized PDFs for the full \textit{Gaia} sample together with the exoplanetary systems in phase space overdensities and in the field. This projection clearly illustrates that the overdensity systems have a lower vertical velocity dispersion than the field systems, despite being confined to the same vertical range relative to the Galactic mid-plane.
}
\end{figure*}
For several of the galactic-dynamical features observed in position-velocity space, it is challenging to determine exactly which physical mechanism triggered its formation. The basic requirement for the generation of these features is the time variation of the gravitational potential, which is easily satisfied given the ubiquitous perturbations that galaxy disks are subjected to, both through hierarchical galaxy formation and secular evolution \citep[e.g.][]{gomez17,antoja18,laporte18,laporte19,fragkoudi19}. Due to the ubiquity and complexity of such perturbations, it is unknown whether the current kpc-scale phase space overdensities are related, and thus were triggered by a single perturbation (or by a small number of them), or if there have been many perturbations that are superimposed and may even form interference patterns. For this reason, it is potentially insightful to attempt a decomposition of 6D phase space to identify planetary systems in overdensities that have a better-constrained origin.

\subsubsection{Selection of planetary systems associated with the\\Gaia phase space spiral}
After the release of \textit{Gaia} DR2, several studies have revealed a `phase space spiral' in the space spanned by the height above the Galactic mid-plane and the vertical velocity ($z{-}v_z$; e.g.\ \citealt{antoja18}). This feature manifests itself independently of the stellar age, which has been used to argue against a bar origin and favor the interpretation that it was generated by a recent disk crossing of the Sagittarius dwarf galaxy \citep{laporte19}. Here we consider the kinematics of the exoplanet host stars in relation to the phase space spiral and investigate whether the architectures of the associated planetary systems differ relative to the other overdensities considered in \S\ref{sec:streams}.

In the left-hand panel of \autoref{fig:zvz}, we show the two-dimensional distribution of the stellar sample in the $z{-}v_z$ plane, together with the planetary systems, colored by their phase space density classification. As in \autoref{fig:rvphi}, we apply an unsharp mask to bring out the contrast relative to the local background, and we use bins with size $\{\delta z,\delta \vz\}=\{12.5~\pc,0.8~\kms\}$. Using an unsharp mask has the advantage of better bringing out the phase-space spiral of \citet{antoja18} in star counts \citep{laporte19}. Stars near the center are unperturbed, whereas stars in the spiral structure are thought to have been perturbed by a passing satellite. This two-dimensional projection is again accompanied by two sets of kernel density-estimated PDFs of the $y$-axis that were constructed analogously to those in \autoref{fig:rvphi}, but here show the distributions of vertical velocities. Each set corresponds to a different maximum height above the mid-plane, with the middle and right-hand panels corresponding to total widths of 0.2~kpc and 1.0~kpc, respectively. These one-dimensional histograms with a maximum height cut again serve the purpose of avoiding dilution by the winding of the phase space spiral in the left-hand panel -- especially the middle panel serves as an attempt to cleanly identify planetary systems associated with the outer parts of the spiral.

\begin{figure*}
\centering
\includegraphics[width=\hsize]{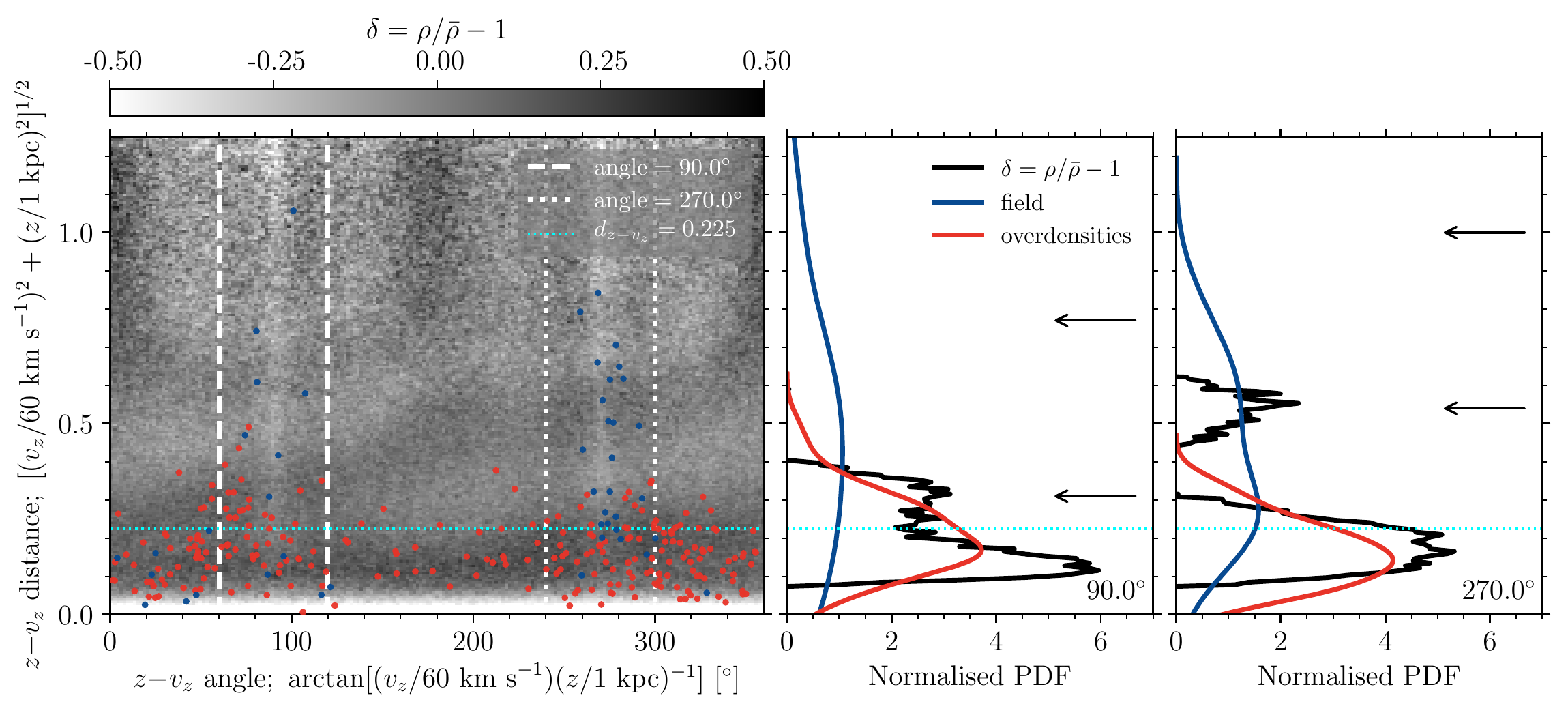}%
\caption{
\label{fig:zad}
Distribution of the phase space spiral from \autoref{fig:zvz} in polar coordinates, which unwinds the spiral by showing the normalized distance from the origin in $z{-}v_z$ space as a function of the azimuthal angle in that space (left). The quantitative formulation of each axis is provided in the axis label, with the normalizations of $z$ and $v_z$ chosen such that the phase space spiral is approximately circular. As in \autoref{fig:zvz}, the gray scale shows a two-dimensional, unsharp-masked histogram. In this projection, the phase space spiral manifests itself as linearly-increasing, dark bands. The horizontal cyan dotted line separates the Galactic mid-plane (below) from perturbed stars (above). The positions of the exoplanet host stars are again indicated by the \{blue, red\} symbols for \{field, overdensity\} systems. The vertical white lines indicate the angles where the Galactic mid-plane is situated (top and bottom in \autoref{fig:zvz}), for which the middle and right-hand panels again show the normalized PDFs for the full \textit{Gaia} sample together with the exoplanetary systems in phase space overdensities and in the field. The arrows mark where the dark band in the left-hand panel cross the angular range marked by the white lines and where the Sagittarius passage is thought to have generated phase space structure. We see that the exoplanetary systems in overdensities indeed show an excess at the first of these loci, suggesting that at least some of the overdensities may have been generated by the cosmological assembly of the Milky Way.
}
\end{figure*}
The most striking result of \autoref{fig:zvz} is the small vertical velocity dispersion of the planetary systems in overdensities ($\sigma_z\approx10~\kms$) compared to the field systems ($\sigma_z\approx20~\kms$), and its close correspondence to the overall vertical velocity dispersion of the overdense ($\delta>0$) stellar sample ($\sigma_z\approx7~\kms$; for reference, the dispersion of the complete data set is $\sigma_z\approx30~\kms$). This vertical velocity dispersion difference arises despite the fact that the exoplanet host stars follow the same vertical spatial distribution around the Galactic mid-plane (see the left-hand panel of \autoref{fig:zvz}). Also note that none of these velocity dispersions are sufficiently high to correspond to the Galactic thick disk, for which $\sigma_z\approx50~\kms$ \citep[e.g.][]{blandhawthorn16}. Instead, they correspond to the young thin disk, as is appropriate for the age cuts that are applied to the planetary system sample (spanning $1{-}4.5$~Gyr).

Upon closer inspection of the vertical velocity distributions, the planetary systems in overdensities show a slight enhancement at velocities of $v_z\approx15~\kms$, which is where the phase space spiral in the left-hand panel connects to the stellar population in the Galactic mid-plane. This suggests that it might be possible to isolate planetary systems that occupy the outer reaches of the phase space feature that was generated by the passage of the Sagittarius dwarf galaxy. To enable a cleaner selection of these systems, we project the phase space spiral in polar coordinates, using a simple parameterization in which the spiral is close to circular. Specifically, we normalize the height above the mid-plane to 1~kpc and the vertical velocity to $60~\kms$, allowing us to define the `distance' in $z{-}v_z$ space relative to the center of the phase space spiral as
\begin{equation}
    \label{eq:dvz}
    d_{z{-}v_z}=\left[\left(\frac{v_z}{60~\kms}\right)^2+\left(\frac{z}{1~\kpc}\right)^2\right]^{1/2} ,
\end{equation}
and the `angle' in $z{-}v_z$ space as
\begin{equation}
    \label{eq:avz}
    \alpha_{z{-}v_z}=\arctan{\left[\left(\frac{v_z}{60~\kms}\right)\left(\frac{z}{1~\kpc}\right)^{-1}\right]} ,
\end{equation}
such that $\alpha_{z{-}v_z}=0\degr$ for $v_z=0$ (stars at their maximum height above the plane) and $z>0$ and $\alpha_{z{-}v_z}=90\degr$ for $v_z>0$ and $z=0$ (stars in the plane moving upward). When unwinding the phase space spiral using the polar coordinates defined in equations~(\ref{eq:dvz}) and~(\ref{eq:avz}), it appears as a set of linear bands in $z{-}v_z$ angle-distance space ($\alpha_{z{-}v_z}{-}d_{z{-}v_z}$), which facilitates a more straightforward association of planetary systems with this structure.

In the left-hand panel of \autoref{fig:zad}, we show the polar projection of the phase space spiral, using the same unsharp masking technique as in \autoref{fig:rvphi} and~\ref{fig:zvz}, and adopting bins of $\{\delta \alpha_{z{-}v_z},\delta d_{z{-}v_z}\}=\{1\fdg8,6.25\times10^{-3}\}$. The dark bands are clearly visible and in this projection are well-described by a linear gradient of ${\rm d}d_{z{-}v_z}/{\rm d}\alpha_{z{-}v_z}\approx 1.3\times10^{-3}/\degr$. Stars near the bottom are unperturbed, whereas stars in the ascending dark-shaded bands are thought to have been perturbed by a passing satellite. Most planetary systems are clustered around small phase space distances near the bottom of the figure, corresponding to low vertical velocities and a position within the Galactic mid-plane. This reflects a discovery bias, because the Sun's position close to the mid-plane means that this is where most exoplanetary systems are discovered. However, the vertical velocity should be unaffected. In these phase space coordinates, planetary systems with high vertical velocities situated in the Galactic mid-plane should reside at $\alpha_{z{-}v_z}=90\degr$ and $\alpha_{z{-}v_z}=270\degr$ for positive and negative vertical velocities, respectively (compare \autoref{fig:zvz}). For a phase space angle range around the first of these values (with a total width of $60\degr$, marked by the white lines in \autoref{fig:zad}), we indeed identify an excess of planetary systems at large phase space distances ($d_{z{-}v_z}>0.225$, chosen to be the local minimum in the distribution of all \textit{Gaia} stars at $\alpha_{z{-}v_z}=90\degr$, see the middle panel) relative to other angles.

The middle and right-hand panels of \autoref{fig:zad} show the one-dimensional distributions of the distance from the center of the phase space spiral at angles of $\alpha_{z{-}v_z}=90\degr$ (stars in the plane moving upward) and $\alpha_{z{-}v_z}=270\degr$ (stars in the plane moving downward), respectively. The arrows indicate where the phase space spiral crosses the selected range of angles in the left-hand panel. For phase space distances $d_{z{-}v_z}<0.6$, the histogram of the full stellar sample indeed shows clear peaks at these loci. These peaks do not appear for larger phase space distances ($d_{z{-}v_z}>0.6$), because the stellar phase space density contrast becomes too small -- when integrating out the phase space angle, the faint overdensity visible in the left-hand panel is therefore cancelled out by the pronounced underdensities within the angular window (the light vertical bands). Most importantly, the middle and right-hand panels show that the planetary systems in overdensities also exhibit an excess at the phase space distance where the phase space spiral first passes through the selected angular window at $\alpha_{z{-}v_z}=90\degr$, whereas the field systems exhibit no excess at the same angle. This implies that some of the planetary systems in overdensities may have been affected by the Galactic disc's response to the passage of the Sagittarius dwarf galaxy. To investigate these systems further, we select all planetary systems with $d_{z{-}v_z}>0.225$. Given that these all reside close to the Galactic mid-plane ($z\approx0$), this range of $z{-}v_z$ distances can be translated into an approximate vertical velocity cut of $v_z\ga13.5~\kms$.

We find that these planetary systems with high vertical velocities do not exhibit any direct correlation with the planar phase space structure in the $R{-}\vphi$ plane of the Galactic disk (considered in \autoref{fig:rvphi}). The high-vertical velocity systems represent a random draw from the field and overdensity populations in that plane -- their $\vphi$ distribution is statistically indistinguishable from the population at large, which includes the planetary systems at small $z{-}v_z$ distances with low $v_z$. This suggests that the phase space spiral identified in $z{-}v_z$ space (\autoref{fig:zvz} and~\ref{fig:zad}) is independent of the ridge-like overdensities identified in $R{-}\vphi$ space (\autoref{fig:rvphi}). This suggestion of independent origins is consistent with the inference of \citet{antoja18} that these features have different dynamical ages. It seems that some of the overdensities considered here indeed may have different physical origins (i.e.\ having been generated by the bar, spiral arms, and/or satellite perturbations) and thus have been able to affect their populations of exoplanetary systems over different timescales. In order to understand exactly how stellar phase space clustering affects planetary system architectures, it is therefore important to look for statistical differences in planet properties between the identified phase space overdensities. As the origin of these dynamical features is better understood, this will also shed new light on the physics driving the link between galactic environment and planetary systems.

\subsubsection{Properties of planetary systems associated\\with the Gaia phase space spiral}
\begin{figure*}
\centering
\includegraphics[width=0.7\hsize]{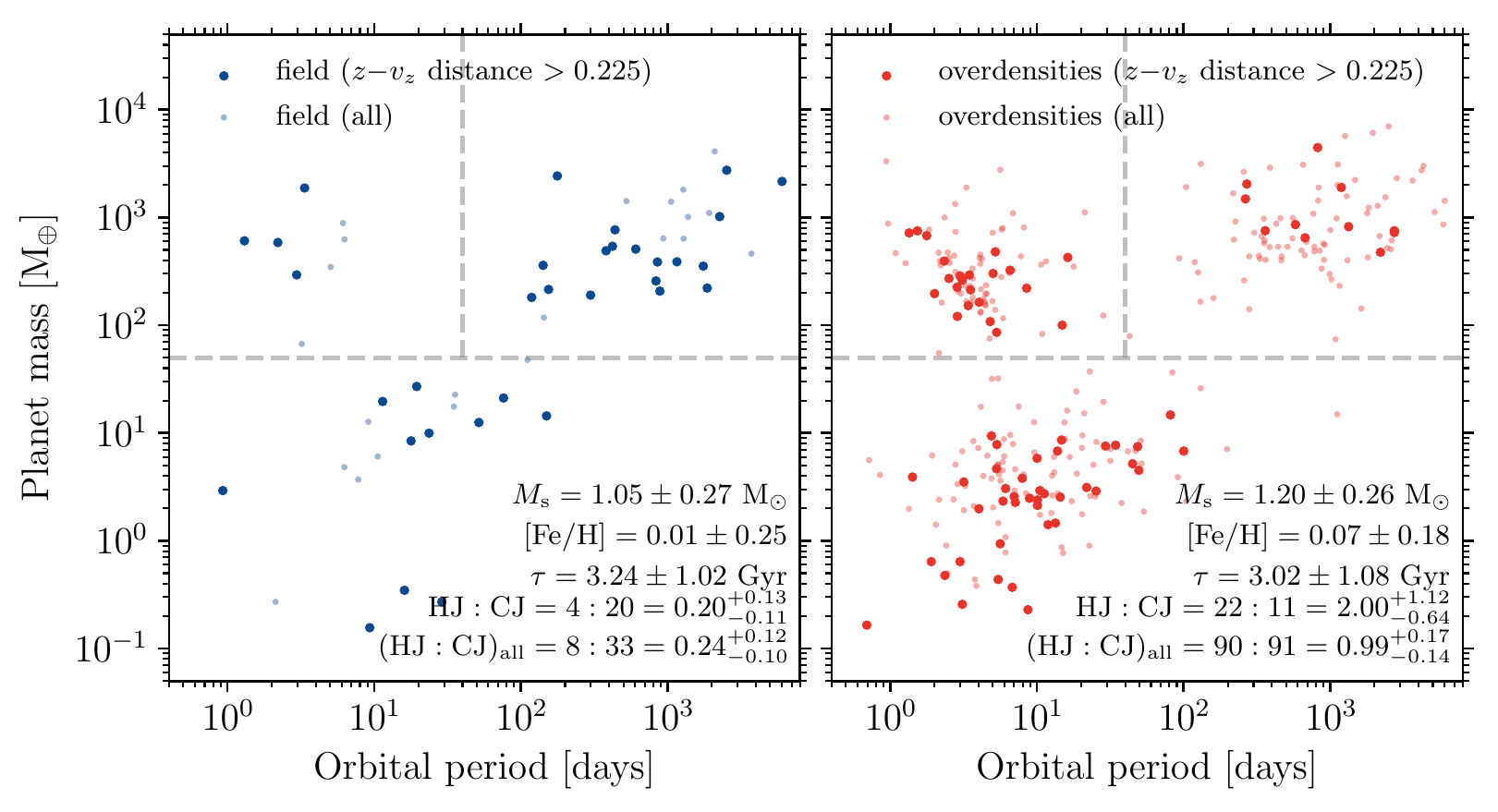}%
\caption{
\label{fig:planets_spiral}
Masses and orbital periods of planets in field (left) and overdensity (right) systems, for all planets (small symbols) and those located at $z{-}v_z$ distances $>0.3$ in \autoref{fig:zad}, corresponding to the outer regions of the phase space spiral in \autoref{fig:zvz} (large symbols). Given that planets are mostly detected close to the Galactic mid-plane, this generally corresponds to planetary systems with high vertical velocities ($v_z\ga20~\kms$). Both panels exhibit the characteristic difference between field and overdensity planets discovered by \citet{wklc20} that motivated this study, but when restricting the sample to systems with high vertical velocities the difference between field and overdensity planets becomes even more pronounced. This means that the planetary system properties do not depend the most strongly on the peculiar velocity of the host star \citep{mustill21}, but on whether or not the host star is part of a phase space overdensity. Finally, the median and standard deviation of the ages of the systems at $z{-}v_z$ distances $>0.225$ (again separating perturbed and unperturbed host stars) are provided in each panel, showing that both classes of system are young, with high-velocity field systems being marginally younger than high-velocity overdensity systems.
}
\end{figure*}
In \autoref{fig:planets_spiral}, we show the distribution of planets in the orbital period-planet mass ($P{-}\mpl$) plane, both for field systems and for systems in overdensities. In both cases, we highlight the planetary systems at large $z{-}v_z$ distances ($d_{z{-}v_z}>0.225$, corresponding to $v_z\ga13.5~\kms$), of which the ones in overdensities are associated with the outer regions of the $z{-}v_z$ phase space spiral. The difference in orbital period distributions between field and overdensity systems is again clearly visible, but even visually the difference seems to be considerably more extreme for the systems with high vertical velocities than for the entire planet sample. As in \autoref{fig:planets_streams}, we consider the hot Jupiter-to-cold Jupiter ratio \hjcj\ as a means of quantifying the difference. Among the high-$d_{z{-}v_z}$ field systems, only 4 out of 24 giant planets are hot Jupiters. By contrast, among the high-$d_{z{-}v_z}$ overdensity systems, 22 out of 33 giant planets are hot Jupiters. In other words, these ratios differ by a factor of $\sim10$. With a statistical significance of $\sim4\sigma$, this is the most extreme difference in planet population properties between an overdensity sample and field sample that has been identified to date (for the full planet sample, \citealt{wklc20} found that the ratios differ by a factor of 3.8), and suggests that the phase space spiral may offer a unique opportunity to help constrain how exactly galactic dynamics shape planetary systems.

A recent study by \citet{mustill21} has suggested that the separation into field and overdensity systems reflects an implicit separation into systems belonging to the thick disk and thin disk, respectively. The key idea is that the phase space clustering found by \citet{wklc20} is dominated by the velocity component rather than the spatial dimensions, such that the field systems have low phase space densities because of their large peculiar velocities, thought to be characteristic of the thick disk. This would be important, because the thick disk is much older than the thin disk, indicating that the separation into field and overdensity systems could be affected by an age bias that may explain differences in hot Jupiter occurrence. \autoref{fig:planets_spiral} now shows that this argument does not hold. Both the field and overdensity samples in that figure have been selected to have the highest vertical velocities of the entire planet sample, yet they exhibit the strongest difference in hot Jupiter occurrence identified to date, with $(\hjcj)_{\rm overdensity}/(\hjcj)_{\rm field}\sim10$. By contrast, when splitting their sample by peculiar velocity, \citet{mustill21} find that the ratios differ by a factor of only $(\hjcj)_{{\rm low}~v}/(\hjcj)_{{\rm high}~v}=1.6$. We infer that planetary system architectures do not depend on whether the host star is moving with a high (vertical) velocity, but on whether it does so \textit{together with other nearby stars}.

The highly significant difference in hot Jupiter occurrence between field and overdensity systems with high vertical velocities warrants a careful assessment of any possible covariance with other quantities that may bias the measurement. Both panels of \autoref{fig:planets_spiral} again list the medians and standard deviations of the distributions of host stellar masses, metallicities, and ages. These are indistinguishable between both samples, indicating that the difference does not reflect any obvious bias in the host stellar properties.\footnote{The close correspondence between the age and metallicity distributions of the field and overdensity systems adds further, independent evidence against the suggestion that the field systems might belong to the thick disk, because the age-metallicity relation of the thick disk differs strongly from the thin disk \citep[e.g.][]{hayden17,buder19}. As a result, there are barely any thick disk stars present at the ages, metallicities, and vertical velocities that characterize the field population in our exoplanet sample \citep[e.g.][]{bovy12,mackereth17}. See \S\ref{sec:perturb} for further discussion of the chemical abundances of exoplanet host stars in phase space overdensities and the field.}

\begin{figure}
\centering
\includegraphics[width=\hsize]{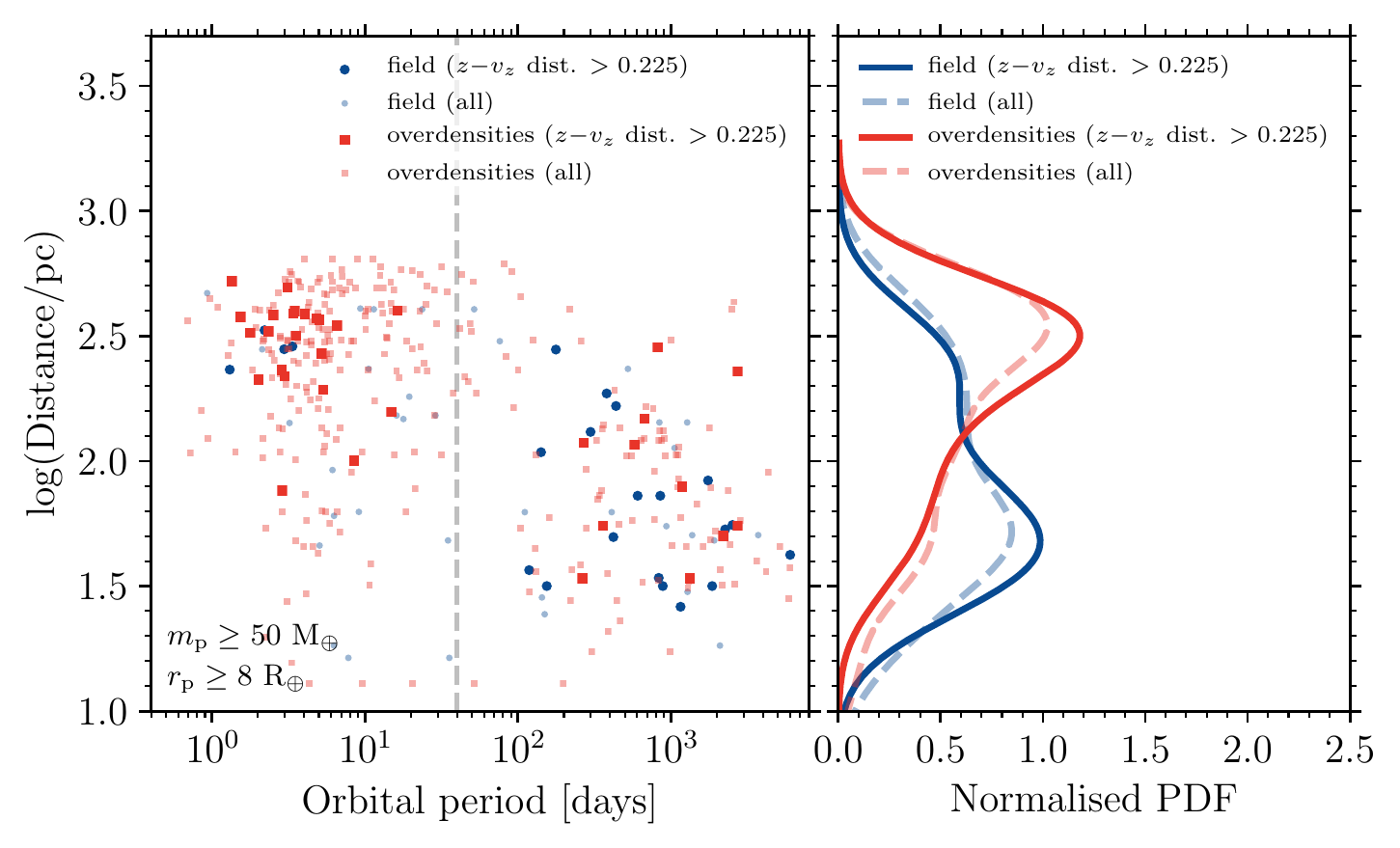}%
\caption{
\label{fig:distance_spiral}
Distance distribution of giant planets in the field and in overdensities, for all planets (small symbols) and those located at $z{-}v_z$ distances $>0.225$ in \autoref{fig:zad} (large symbols), shown as a function of orbital period (left) and as a kernel density estimate-smoothed histogram (right). The distance distributions of all sub-samples overlap, but their shapes differ. Importantly, when restricting the sub-samples to $d\leq300~\pc$, the distance distributions are statistically indistinguishable (see the text), but the strong difference in the hot Jupiter-to-cold Jupiter ratio (\hjcj) from \autoref{fig:planets_spiral} persists. This demonstrates that the difference in \hjcj\ does not result from a distance bias.
}
\end{figure}

As in \S\ref{sec:streams}, \autoref{fig:distance_spiral} shows the distribution of the giant planets in the orbital period-distance ($P{-}d$) plane, this time for each of the samples shown in \autoref{fig:planets_spiral}. The distance distributions again span a similar, overlapping range, but the shapes of the distributions differ. As before, the field systems tend to reside at closer distances than the overdensities, which persists when restricting the sample to high-$v_z$ systems with $d_{z{-}v_z}>0.225$. Given that the detectability of cold Jupiters is more challenging, this could potentially explain our observation in \autoref{fig:planets_spiral} that \hjcj\ is significantly higher in the high-$v_z$ overdensity systems. However, when restricting the sample to planetary systems at distances $d\leq300~\pc$ as in \citet{wklc20}, the distance distributions are statistically indistinguishable (with a KS test $p$-value of $p_{\rm KS}=0.13$, as opposed to $p_{\rm KS}=7.4\times10^{-3}$ for the full sample) and the left-hand panel of \autoref{fig:distance_spiral} shows that the major difference in \hjcj\ remains, despite a non-negligible decrease of the number of planetary systems due to the distance cut. Quantitatively, for the distance-restricted sample we find $(\hjcj)_{\rm field}=0.15^{+0.13}_{-0.10}$ and $(\hjcj)_{\rm overdensity}=0.73^{+0.38}_{-0.27}$, corresponding to $(\hjcj)_{\rm overdensity}/(\hjcj)_{\rm field}\sim5$. This means that a distance bias alone cannot be responsible for the highly significant excess of hot Jupiters in planetary systems associated with the the $z{-}v_z$ phase space spiral.

Summarising the above discussion, we find that the planetary systems associated with the phase space feature that was likely generated by a recent passage of the Sagittarius dwarf galaxy have a hot Jupiter occurrence ($\hjcj=2.00^{+1.12}_{-0.64}$) that is an order of magnitude higher than for field systems at similar vertical velocities ($\hjcj=0.20^{+0.13}_{-0.11}$). This is the largest difference in planet properties found to date between field and overdensity sub-samples, and shows that peculiar velocity alone is a poor predictor of planetary system architectures. Instead, planetary system architectures depend on whether the host star is moving together with other stars. We discuss the implications of these results further in \S\ref{sec:disc}. Finally, as mentioned in \S\ref{sec:streams}, these findings should be treated as indications to prompt future work, rather than definitive proof. A quantitative assessment of hot Jupiter occurrence requires accounting for potential differences in detectability, specifically folding in the non-detection rates of the different surveys and detection methods. This is an important area of interest for follow-up studies.

\section{Discussion}
\label{sec:disc}
The association of the phase space overdensities around exoplanetary systems with the kpc-scale phase space corrugations of the Galactic disk unambiguously demonstrates that the overdensities are not the remains of the birth clusters in which the planetary systems might have formed. The definitive proof would be to measure the chemical abundance spreads of the stellar population contained in each overdensity. Stellar clusters and associations originate from the same parent molecular cloud, causing them to have metallicity spreads smaller than the measurement uncertainty \citep[up to a few 0.01~dex, e.g.][]{krause20,casamiquela21}. It seems inevitable that the metallicity spreads of the overdensities considered here are larger. The metallicity spreads of exoplanet host stars within each of the four specific overdensities in \autoref{fig:planets_streams} ($R{-}\vphi$ ridges) and \autoref{fig:planets_spiral} (phase space spiral) are $0.1{-}0.2$~dex, much larger than those within a single molecular cloud \citep[$<0.05$~dex; e.g.][]{esteban18,kreckel19,kreckel20,casamiquela21}. While a clear picture is now emerging of the physical nature of the phase space overdensities, the results of this paper also raises several immediate follow-up questions that need to be answered in order to understand how exactly the galactic environment affects the properties of planetary systems. We briefly discuss these here.

\subsection{What are the lifecycles of the phase space overdensities?}
\label{sec:lifecycle}
There is a rapidly-growing literature that discusses the physical origin and nature of the complex phase space structure of the Galactic disk, and does so in much more detail than the simple terms (birth cluster versus galactic dynamics) that the discussion in this paper necessarily focused on thus far. It has been demonstrated at length that the kpc-scale phase space overdensities can be generated by any form of perturbation in the gravitational potential, be it spiral arms \citep{hunt18,quillen18}, the Galactic bar \citep{dehnen00,fragkoudi19,monari19,laporte20}, or satellite galaxy passages \citep{minchev09,antoja18,laporte19,hunt21}. The lifetimes of the overdensities differ between these mechanisms, and they each have a unique imprint on the phase space structure of the Galactic disk.

Galaxy haloes consist of substructure \citep[e.g.][]{springel08}, and satellite galaxy passages
are effective at generating bending waves, which create phase space spirals in the $z{-}v_z$ plane (\autoref{fig:zvz}), alongside ripples in $R{-}\vphi$ space \citep{chequers18,laporte19,bennett21,blandhawthorn21,hunt21}. By contrast, resonances driven by the bar or spiral arms generate finely-spaced ridges in the $R{-}\vphi$ plane (\autoref{fig:rvphi}), but these are few in number because each ridge is uniquely connected to a single resonance \citep[e.g.][]{antoja18}. Resonance-driven overdensities are expected to conserve their vertical angular momenta \citep[modulo some minor deviations, see e.g.][]{monari19}, but \citet{ramos18} observe that this does not apply to the Sirius overdensity, whereas it does to the Hercules overdensity \citep[which indeed coincides with a bar-driven resonance, e.g.][]{dehnen00,monari19,chiba21}. These arguments strongly suggest that the complex phase space structure of the Galactic disk is generated by multiple physical mechanisms acting simultaneously and interfering with one another \citep[e.g.][]{hunt18,hunt21}.

It would not be surprising if the various phase space overdensities in the solar neighborhood do indeed have a variety of physical origins. The Milky Way has a bar and spiral arms \citep[e.g.][]{blandhawthorn16}, and the Galactic disk has been perturbed by satellite galaxy accretion throughout its history \citep[e.g.][]{pricewhelan15,bergemann18,helmi20,kruijssen20c,laporte20b}. These different sources of perturbation are potentially highly beneficial for understanding the correlation between planetary system properties and the ambient stellar phase space density. After all, different physical origins would imply that the overdensities also have different lifetimes and evolutionary lifecycles, plausibly resulting in corresponding differences in how the planetary systems have been affected.

The bar is thought to have an age of $\sim8$~Gyr \citep[e.g.][]{bovy19,grady20}, implying that it has been generating phase space overdensities for a similar duration. Phase space overdensities generated by satellite galaxy perturbations live for several Gyr in numerical models \citep[e.g.][]{laporte19,hunt21}. Irrespectively of the specific mechanism, we conclude that the phase space overdensities that are correlated with planetary system architectures have lifetimes of at least several Gyr. This is reassuring, because a lifetime similar to the planetary system age (here $1{-}4.5$~Gyr) is required to explain why planetary system properties are correlated with the \textit{current} phase space density -- otherwise, the current association of a planetary system with a phase space overdensity might not be representative for its evolutionary history.

Now that we can begin to associate planetary systems with specific overdensities as demonstrated in this paper, constraining the detailed origin and evolutionary lifecycle of each overdensity would help explain the differences that we observe. For instance, if the Hercules overdensity corresponds to a resonance and Sirius to a satellite galaxy perturbation, do these overdensities have different ages and lifecycles, and could this help explain the difference in their hot Jupiter-to-cold Jupiter ratio (\hjcj, see \autoref{fig:planets_streams})? This would certainly be possible, because the Galactic bar is thought to be a long-lived feature, whereas the satellite passage that may have generated Sirius potentially occurred more recently.

Similarly, we discussed above that the phase space spiral is likely generated by a satellite perturbation. Based on the difference in \hjcj\ between the planetary systems associated with Sirius (\autoref{fig:planets_streams}) and with the phase space spiral (\autoref{fig:planets_spiral}) despite indistinguishable distributions of host stellar masses, metallicities, ages, and distances, we propose that these two overdensities may have been generated by different satellite passages occurring at different times. Given that \hjcj\ is higher for the systems associated with the phase space spiral (which is a sign of perturbation), the perturbation that triggered Sirius may have occurred more recently than the perturbation that triggered the phase space spiral. These are testable predictions -- due to phase mixing, the separation between phase space corrugations generated by a single perturbation shrinks with time \citep[e.g.][]{minchev09,gomez12}, implying that a Fourier analysis of phase space ripples (such as those in \autoref{fig:rvphi}) could potentially be used to age-date individual overdensities.\footnote{This is complicated by the interference from bar resonances \citep[e.g.][]{monari19,laporte20}. It might be possible to address complication by characterising low-angular momentum groups within the Galactic co-rotation radius, which would allow a robust characterization of the separation between ridges without any interference from bar-related structures.} Conversely, we speculate that it might eventually be possible to age-date overdensities using their \hjcj\ ratios.

\subsection{How much time do planetary systems spend in phase space overdensities?}
\label{sec:time}
Once it is established what the lifecycle of a particular phase space overdensity has been, i.e.\ how it was generated, what its age is, and how it evolved, the next question is how long the associated planetary systems have been part of that overdensity. The time spent in the overdensity is a crucial quantity to determine how an overdensity might have affected the properties of its associated planetary systems. To address this question, it is first important to emphasize that the phase space overdensities are moving groups, which means that their constituent stars remain in the parent overdensity until phase mixing removes them from the structure. In particular, bar-driven overdensities are thought to migrate outwards through the Galactic disk \citep[e.g.][]{chiba21,hunt21}. While the stars can remain in an overdensity for a long time, the overdensity itself might not have formed at its current location.

The key question is then whether the overdensities originated before or after the formation of their associated planetary systems. Undulations like the phase space structures considered here are also observed in the interstellar medium of the solar neighbourhood \citep{alves20}, and analogous structures may be observed in external galaxies \citep{henshaw20,gomez21}. Therefore, it is possible that the same perturbations that generate overdensities in the stars also do so in the gas, potentially even triggering or shutting down star formation \citep[e.g.][]{laporte20b,ruizlara20}. The current phase space clustering of the stars is then simply inherited from the phase space structure of the interstellar medium. As discussed in \S\ref{sec:perturb}, this scenario does have some difficulties in explaining the observed correlation between planetary system properties and phase space density. Nonetheless, if the phase space structure is present at birth, then the planetary systems spend their entire lives within the overdensities, unless phase mixing erases them on a timescale shorter than their ages. This process might explain why the fraction of exoplanetary systems in overdensities drops precipitously at ages $>5$~Gyr \citep[see Extended Data Figure~10 of][]{wklc20}.

Alternatively, the overdensities may be generated after the formation of the associated planetary systems. This plausibly applies to the phase space spiral, which can be traced in the stellar population at all ages up to 9~Gyr and provides clear evidence that the entire disk has responded to a perturbation that may have occurred as recently as $1{-}2$~Gyr ago \citep{laporte19}. In this case, it seems inevitable that some of the older associated planetary systems may have formed well before the overdensity originated. This may apply to other overdensities too -- the age distributions of planetary systems within overdensities are quite flat between $1{-}4.5$~Gyr (see \S\ref{sec:perturb}), meaning that any overdensity that was generated during that time interval will contain planetary systems that formed before and after the overdensity itself.

In summary, it appears that the relative ages of overdensities and their associated planetary systems may vary considerably, both between different overdensities and within them. However, once an overdensity has been generated, its members remain associated until phase mixing disperses the overdensity. This makes it a priority to determine the ages of the overdensities; in conjunction with the planetary system ages, this will allow us to determine precisely which planetary systems have resided in their parent overdensity since birth, and how much time any of the planetary systems have spent in their current phase space overdensities.

\subsection{Which mechanisms perturb(ed) the planetary systems in phase space overdensities?}
\label{sec:perturb}
Empirically, it is becoming well-established that the properties of planetary systems depend on the degree of ambient stellar clustering in position-velocity phase space. This implies that some form of external perturbation is capable of affecting planetary systems. However, attempts at identifying the physical mechanisms responsible have been obstructed by the uncertain physical nature of the phase space overdensities (see the discussion in \S\ref{sec:intro}). We have now demonstrated that the overdensities do not represent the remains of the planetary system's birth cluster, but the way in which the Galactic disk responds to perturbations.\footnote{The specific source of the perturbation does not need to be known, because the disk's response to perturbations is set by its vertical frequency and is sustained by its self-gravity, independently of the nature of the perturber \citep[e.g.][]{darling19}.} Therefore, we can reassess this question.

In brief, there are at least three different ways in which phase space clustering may affect planetary systems.
\begin{enumerate}
    \item Stars in overdensities and in the field formed in different locations, of which the phase space density is an indirect tracer. This may be at different galactocentric radii, such that both populations have experienced different degrees of radial migration. Alternatively, if the overdensity has been generated by a satellite galaxy passage, the constituent stars potentially may have originated in the satellite galaxy itself. Because exoplanets are detected close to the solar circle, we do not know how planet demographics may depend on the birth location within the Milky Way or its satellites.
    \item There is a causal relationship between the \textit{generation} of the phase space overdensity and the perturbation of the planetary system. This is the case if the overdensities represent the remnants of the birth stellar clusters, in which external photoevaporation of the protoplanetary disk or stellar encounters may have affected the formation or early evolution of the planetary systems.
    \item There is a causal relationship between \textit{membership} of a phase space overdensity and the perturbation of the planetary system. In this case, the membership of a phase space overdensity somehow alters the dynamical history of a planetary system, potentially by stellar encounters or by the tidal interaction with (substructure in) the Galactic disk, which slowly changes the planetary system properties over time.
\end{enumerate}
The first two of these reflect changes that are imprinted at birth or through covariance with some other natal property. By contrast, the third option describes a situation in which a planetary system's membership of a phase space overdensity continues to perturb the system over long timescales. Contrary to the first two scenarios, this predicts that the planetary system demographics in overdensities depend on host stellar age or on the time spent in the overdensity (whichever is shortest).

To first order, our finding that the overdensities have been generated by galactic-dynamical processes make it more likely that the perturbation of planetary systems takes place during their evolution rather than at formation (although both might still apply). After all, the phase space structure can be generated at any time, because at least some phase space overdensities with ages of a few Gyr contain stars that are much older \citep{laporte19}. Similarly, the considerable age and metallicity spreads of the exoplanet host stars associated with individual overdensities suggest that either these overdensities are so old that they pre-date their perturbed planetary systems (in which case the perturbation may take place during planet formation and early evolution), or the perturbation of the planetary systems can take place at any time after their formation. The fact that the ages estimated for some of the overdensities \citep[few Gyr,][]{laporte19,hunt21} are lower than our upper age cut for the planetary systems (4.5~Gyr) strongly favors late-time (dynamical) perturbations over early (radiative or cluster-based dynamical) perturbations.

Nonetheless, it might be that the location of an exoplanetary system in an overdensity is somehow covariant with having formed in a dense and perturbative, clustered environment, such that stellar clustering at birth might still be responsible for the differences between planets currently residing in overdensities and in the field. For instance, the galactic conditions leading to strongly clustered star formation might also favor the resulting planetary systems to end up in phase space overdensities at the present day. However, for clustering at birth to then be responsible for the observed correlation between planetary system properties and current clustering, few (or none of) such systems may end up in the field. This seems infeasible, because there is no known dynamical process that preferentially deposits stars formed in clusters in galactic-scale phase space overdensities. The only way in which this may be possible is if the phase space overdensities are birth features themselves, such as the corrugations found in the interstellar medium of the Milky Way \citep{alves20,henshaw20} and external galaxies \citep{matthews08,elmegreen18,henshaw20,narayan20}, or if they have migrated from their birth sites at smaller galactocentric radii. In either of these cases, the star formation within these birth environments must have been considerably more clustered (and therefore more perturbative to planetary systems) than elsewhere in the disk.

While the formation of the most massive (and correspondingly rare) clusters may indeed be enhanced near resonances \citep[e.g.][]{herrera20} and in spiral shocks \citep[e.g.][]{elmegreen18}, massive clusters are observed to form throughout the disks of the Milky Way \citep[e.g.][]{portegieszwart10,longmore14,krumholz19} and other nearby galaxies \citep[e.g.][]{adamo15,adamo20}. This implies that their potentially disruptive effect during planet formation and early evolution should not be restricted to systems ending up in late-time overdensities. Additionally, massive clusters represent only a few percent of the star formation in Milky Way-like galaxy disks \citep[e.g.][]{kruijssen12d,adamo20}, in strong contrast with the prevalence of the phase space overdensities around exoplanetary systems. Taken together, this suggests that the clear bimodality in several planet properties (e.g.\ the paucity of hot Jupiters and super-Earths in field systems found by \citealt{wklc20} and \citealt{kruijssen20d}, respectively) is not exclusively a relic of the birth environment, and instead supports the interpretation that such environmental dependences may predominantly arise at a later age.

A weaker version of the idea that planetary systems in overdensities were perturbed by their birth cluster is that the overdensity systems represent a different component of the Galactic disk than the field systems, and that this might reflect a more general difference in ambient environment. \citet{mustill21} propose that the overdensity population corresponds to the thin disk, whereas the field population corresponds to the thick disk, with the key distinguishing feature being the magnitude of the peculiar velocity. As we discuss in \S\ref{sec:spiral}, this solution is quite unlikely for two reasons. First, the field systems have age and metallicity distributions that are statistically indistinguishable from those of the overdensity systems, and these distributions are not appropriate for thick disk stars \citep[e.g.][]{bovy12,mackereth17}. Secondly, the sub-samples of field and overdensity systems with high vertical velocities exhibit the same qualitative differences in planet properties as the parent sample, implying that the planetary system architecture does not depend on whether it is moving with a high velocity that potentially indicates thick disk membership, but on whether it is moving together with other stars.

\begin{figure}
\centering
\includegraphics[width=\hsize]{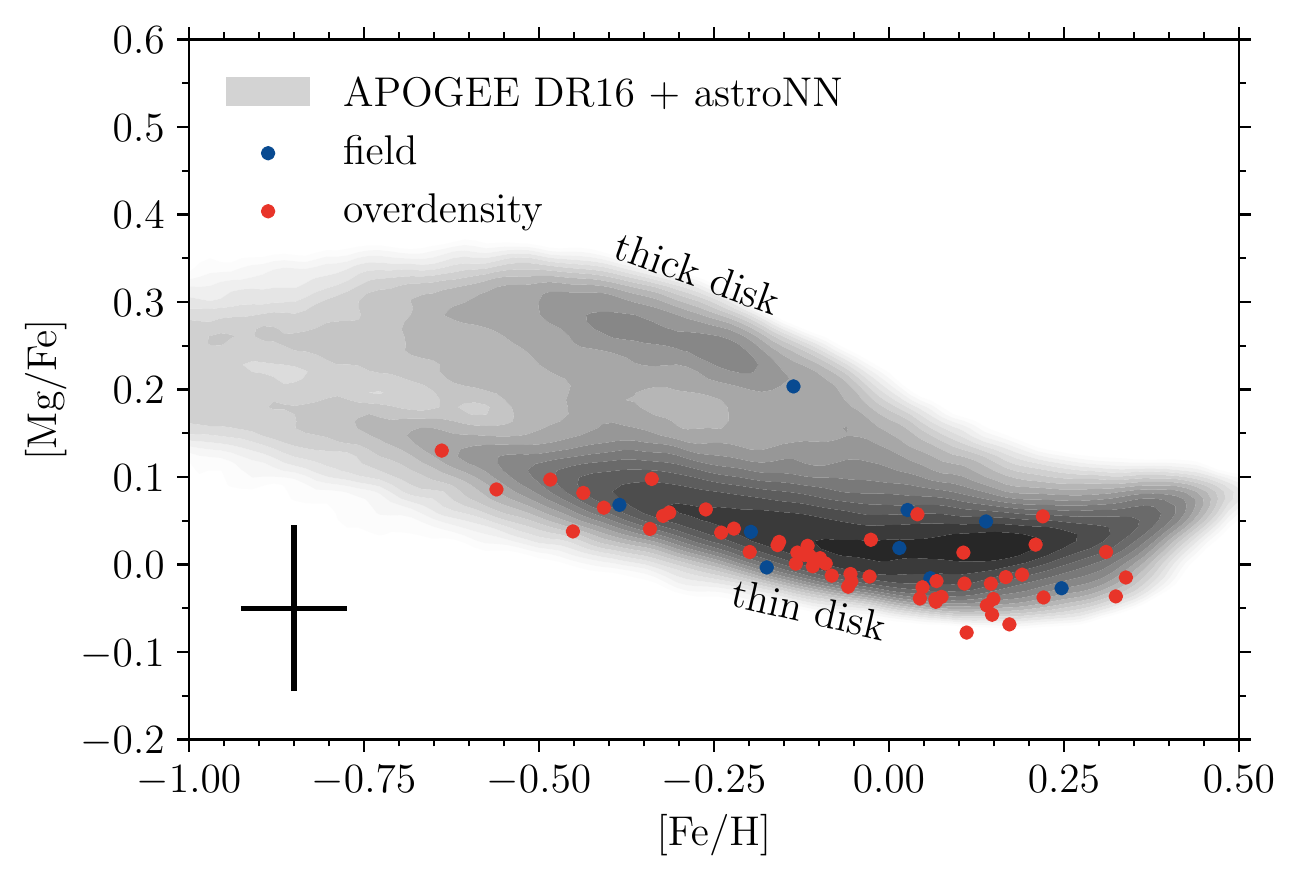}%
\caption{
\label{fig:mgfe}
Distribution of host stars in the \feh--\mgfe\ plane, obtained by cross-matching our planetary system sample to the astroNN catalogue \citep{leung19} from APOGEE DR16 \citep{jonsson20}. Blue symbols represent systems in the field, whereas red symbols indicate systems in overdensities. The gray-shaded contours indicate the distribution of stars in the entire astroNN catalogue, in which the thin and thick disks are clearly visible (see annotations). Typical uncertainties are indicated by the bottom-left error bar. The figure shows that the vast majority of the field and overdensity systems in our sample belong to the young, thin disk, implying that the differences in planet demographics do not result from having formed in different components of the Galactic disk.
}
\end{figure}

Perhaps the most straightforward way of distinguishing between the thin and thick Galactic disks is by considering the chemical abundances of the exoplanet host stars in our sample. At fixed \feh, the thick disk is characterized by elevated $\alpha$-element abundances relative to the thin disk \citep[e.g.][]{blandhawthorn16,hayden17,buder19}. \autoref{fig:mgfe} shows the distribution of host stars in the \feh--\mgfe\ plane, obtained by cross-matching our planetary system sample to the astroNN catalogue of \citet{leung19}, which is derived from APOGEE DR16 \citep{jonsson20}. The cross-match includes 62 planetary systems with unambiguous phase space density classifications (53 in overdensities and 9 in the field). As \autoref{fig:mgfe} shows, only one planetary system belongs to the thick disk. For the age range of our sample ($1{-}4.5$~Gyr), the vast majority of both field and overdensity systems belong to the young, thin disk. This means that the differences in planet demographics between overdensities and the field cannot be explained by the idea that both groups of planetary systems may have formed in different components of the Galactic disk.

Finally, even the favored option that planetary systems have been perturbed at a later age while residing in the current phase space overdensities (most plausibly dynamically) has difficulties. Due to the much lower stellar density than in gravitationally-bound stellar clusters, the stellar encounter rate in the Galactic mid-plane is extremely low. Over 5~Gyr timescales, stellar encounters with a solar system-like planetary system can excite orbital eccentricities of only $e=0.01{-}0.1$ \citep{zakamska04}. However, given the long timescales involved, this may be sufficient to eventually destabilize the planetary system and modify the hot Jupiter-to-cold Jupiter ratio that we use as a diagnostic in this paper. Alternatively, the external tidal field generated by the interstellar medium or the Galactic disk may excite perturbations within planetary systems over Gyr timescales \citep[e.g.][]{kaib13}. As discussed by \citet{kruijssen20d}, \citet{chevance21}, and \citet{longmore21}, any of these long-term dynamical perturbations could also be an efficient mechanism for explaining why overdensities have an elevated occurrence of super-Earths relative to sub-Neptunes, as well as enhanced radius uniformity within multiple systems, and an excess of single-planet systems.

In summary, our findings suggest that the dependence of planetary system properties on ambient stellar clustering is caused by late-time (dynamical) perturbations rather than radiative or dynamical perturbations within the birth cluster. This favors the third of the three options described at the beginning of this subsection as the root of the correlation between phase space density and planetary system properties. However, the impact of the birth environment is not ruled out altogether. Observations of nearby star-forming regions clearly demonstrate that nearby stars can drive protoplanetary disk dispersal through radiative and dynamical perturbations. It remains to be demonstrated how these early mechanisms may impact the long-term demographics of planetary systems. Future work aiming to make this connection should account for the fact that, under solar neighborhood conditions, only a small minority of stars and planetary systems ($5{-}10\%$) forms in gravitationally-bound clusters, and an even smaller fraction ($2{-}3\%$) is born in massive clusters of $M\geq10^4~\msun$ \citep{kruijssen12d,adamo20}.

\subsection{What is the relation between planetary system properties, phase space overdensity, and stellar age?}
\label{sec:age}

The most conclusive way of distinguishing between the scenarios outlined in \S\ref{sec:perturb} is by using the host stellar age information. If the external perturbation of the planetary systems takes place during or shortly after the time of their formation, then \hjcj\ may be constant with age, or even decrease due to the tidal inspiral of hot Jupiters.\footnote{Indeed, the paucity of hot Jupiters in field systems is one of the main reasons why \citet{mustill21} suggest that field systems might belong to the thick disk and have older ages than overdensity systems.} Alternatively, if the external perturbations take place over Gyr timescales and are generated by the phase space overdensities within which planetary systems are currently observed to reside, then \hjcj\ should increase with age. While stellar ages remain notoriously uncertain (and we are currently revisiting and updating these for the planetary system sample considered here; S.~N.~Longmore et al.~in preparation), our identification of individual overdensities now does enable us to make a first attempt at considering the planetary demographics as a function of the stellar age distribution in the parent phase space overdensity.

\begin{figure}
\centering
\includegraphics[width=\hsize]{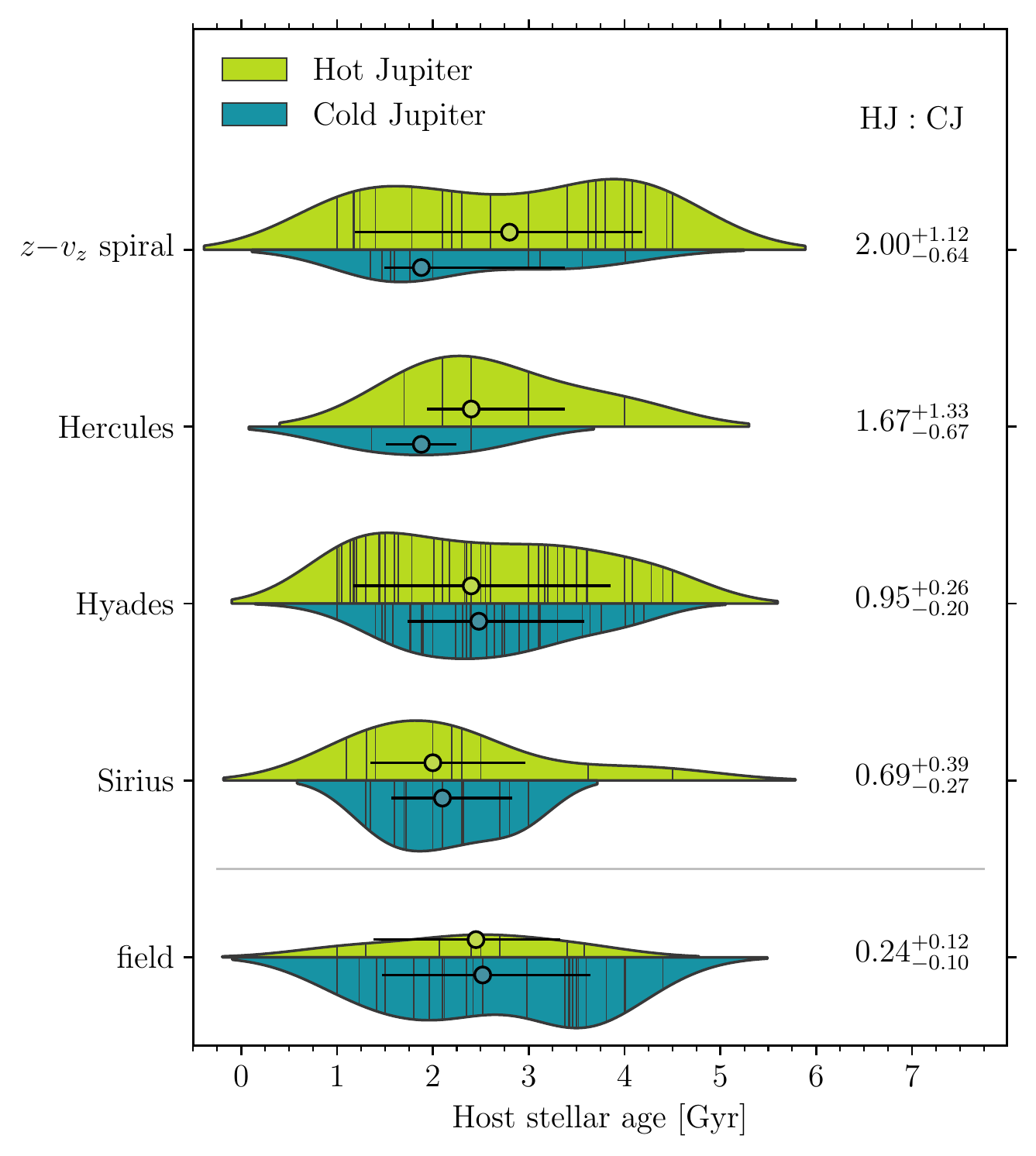}%
\caption{
\label{fig:ages}
Age distribution of host stars in planetary systems hosting giant planets, divided by their membership of individual phase space structures (rows; ordered from bottom to top by increasing hot Jupiter-to-cold Jupiter ratio \hjcj, as quantified on the right) and by planet type (violin sides; light and dark shading refers to hot and cold Jupiters, respectively). Circles with error bars highlight the medians and 16$^{\rm th}$-to-84$^{\rm th}$ percentiles of the distributions, and vertical lines indicate individual data points. The figure shows that \hjcj\ gently increases with host stellar age, both between overdensities that contain stars of different ages (bottom to top) and within each individual overdensity as a function of age (left to right), implying that the correlation between planetary system properties and stellar phase space clustering grows over Gyr timescales.
}
\end{figure}

\autoref{fig:ages} shows the distribution of host stellar ages (taken from the compilation in the \citealt{exoarchive}) for planetary systems hosting giant planets, separated by individual overdensity membership (field, Sirius, Hyades, Hercules, and $z{-}v_z$ spiral) and by planet type (hot and cold Jupiters). The overdensity membership is determined using the same criteria as in \autoref{fig:planets_streams} and \autoref{fig:planets_spiral}. Between the three different overdensities, we see that the hot Jupiter-to-cold Jupiter ratio increases with the median age of the host star. Even when only considering systems within an individual overdensity, such as the Hyades or the phase space spiral (which have the largest sample sizes among the four overdensities and therefore provide the best statistics), we see that \hjcj\ (the ratio between the top and bottom histograms) increases with age above the median (i.e.\ to the right of the upper data point in each overdensity).

Taken together, these trends between \hjcj\ and host stellar age provide tentative evidence that, within overdensities, hot Jupiter production proceeds at a higher rate than hot Jupiter destruction, causing \hjcj\ to increase with host stellar age over Gyr timescales. We do not find any evidence that \hjcj\ may decrease with age. This falsifies the hypothesis that the observed correlation between planet properties and stellar phase space clustering results from hot Jupiter attrition by tidal inspiral towards older ages in field systems \citep{mustill21}. Instead, \autoref{fig:ages} suggests that the opposite holds -- the environmental perturbations that generate the relation between planetary systems and stellar clustering continue to act on overdensity systems long after the planet formation process has ceased. This also suggests that planetary population synthesis models aiming to reproduce the observed exoplanet population \citep[e.g.][]{emsenhuber20} need to account for the disruptive impact of the large-scale galactic environment on planetary systems well beyond the timescale of protoplanetary disk dispersal. This is a major challenge, because it is currently still unknown which physical mechanisms drive this long-term impact of the galactic environment on planetary systems. Solving this problem will require a coordinated effort of all communities involved, linking studies of planet formation and evolution, star formation, and galaxy formation and evolution.

\section{Conclusions}
Following on our recent discovery that the architectures of planetary systems depend on the ambient degree of stellar clustering in position-velocity phase space \citep{wklc20}, we have investigated in this paper what the physical nature is of the phase space overdensities. The goal of this effort is to define the framework needed to determine in the future which physical mechanisms cause the environmental dependence of the planet properties. We use the full catalogue of stars in \textit{Gaia} DR2 with 6D phase space information to characterize the kinematic structure of the Galactic disk and compare it to the properties of planetary systems with phase space classifications. Our main results are as follows.
\begin{enumerate}
    \item
    The phase space overdensities around planetary systems correspond to the well-known, kpc-scale ripples and streams in the Galactic disk that are thought to be generated by resonances from the Galactic bar and spiral arms, as well as by perturbations from passing satellite galaxies. The overdensities do not represent the remains of the planetary systems' birth clusters. (\S\ref{sec:streams})
    \item
    By selecting planetary systems belonging to individual phase space overdensities, we find indications that the planet demographics may differ between overdensities, which potentially have differing physical origins and histories. Specifically, the hot Jupiter-to-cold Jupiter number ratios in the Sirius, Hyades, and Hercules overdensities are $(\hjcj)_{\rm Sirius}=0.69^{+0.39}_{-0.27}$, $(\hjcj)_{\rm Hyades}=0.95^{+0.26}_{-0.20}$, and $(\hjcj)_{\rm Hercules}=1.67^{+1.33}_{-0.67}$. For the fast-moving stars associated with the \textit{Gaia} phase space spiral overdensity, which is thought to have been generated by a passage of the Sagittarius dwarf galaxy, we obtain $(\hjcj)_{\rm spiral}=2.00^{+1.12}_{-0.64}$. All of these ratios are considerably higher than for planetary systems in the field, which have $(\hjcj)_{\rm field}=0.24^{+0.12}_{-0.10}$. In view of these differences, we expect that being able to separate planetary systems by individual overdensities will provide an important diagnostic for identifying the physical mechanisms responsible for perturbing planetary systems. Our findings should be followed up by a thorough detectability assessment for each different overdensity, also folding in the non-detection rates of the different surveys and detection methods. (\S\ref{sec:streams} and \S\ref{sec:spiral})
    \item
    When selecting planetary systems with high vertical velocities ($v_z\ga13.5~\kms$) in the phase space spiral overdensity, the hot Jupiter-to-cold Jupiter ratio is a factor of $\sim10$ higher than among field stars with similarly high vertical velocities. This means that planetary system architectures do not depend on whether the host star is moving with a high (vertical) velocity, but on whether it does so together with other stars. By combining this finding with the indistinguishable age distributions and chemical abundances of these field and overdensity systems, we can rule out the recent suggestion that the field systems belong to the thick Galactic disk. Instead, both the field and overdensity systems are part of the young thin disk. (\S\ref{sec:spiral} and \S\ref{sec:perturb})
    \item
    Thanks to the correspondence of the phase space overdensities around planetary systems to the \textit{Gaia} kpc-scale streams, we can infer that the overdensities are long-lived, and may host their associated planetary systems for several Gyr. We suggest that differences in overdensity ages may explain the variations in planetary system architectures that we find between individual overdensities, and speculate that it may eventually be possible to age-date phase space overdensities using their planet demographics. (\S\ref{sec:lifecycle})
    \item
    Some planetary systems may have formed in overdensities, whereas others may pre-date the formation of the overdensity that they currently reside in. These systems can be distinguished by obtaining accurate ages of exoplanet host stars and their parent overdensities. Doing so will help inform when planetary systems are perturbed by their environments and through which physical mechanisms. (\S\ref{sec:time})
    \item
    Our finding that the correlation of planetary system architectures with ambient stellar clustering reflects a dependence on kpc-scale galactic-dynamical features has made it more likely that this correlation is caused by the late-time (dynamical) perturbations of planetary systems (e.g.\ by stellar encounters or galactic tidal perturbations). However, we do not rule out the possibility that radiative perturbations during planet formation may also have affected planetary systems in phase space overdensities. In this case, the overdensities may be the relics of kpc-scale corrugations and compressions in the interstellar medium that seeded star and planet formation. Perturbative mechanisms relying on the gravitational boundedness of the birth cluster (e.g.\ dynamical interactions between stars within the birth environment) are unlikely to explain the correlation due to the rarity of bound clusters under solar neighborhood conditions. (\S\ref{sec:perturb})
    \item
    Using the currently available stellar ages, we find that the hot Jupiter-to-cold Jupiter ratio within overdensities may increase with host stellar age over Gyr timescales. This suggests that hot Jupiter production proceeds at a higher rate than hot Jupiter destruction in these systems, and that the environmental perturbations that drive the correlation between planetary system properties and stellar clustering continue to act long after the planet formation process has ceased. We propose that accounting for the impact of the large-scale galactic environment on planetary systems long after protoplanetary disk dispersal represents an important direction for improving current planetary population synthesis models. (\S\ref{sec:age})\vspace{2mm}
\end{enumerate}
Our results imply that planetary systems are not just affected by stellar clustering in their immediate surroundings, but by galaxy-scale processes throughout their evolution. It is a major challenge to obtain a complete census of the physical processes at play, which requires combining studies of planet formation and evolution, star (cluster) formation, galactic dynamics, and galaxy formation and evolution. As a first step, we have identified the three most pressing questions in \S\ref{sec:disc}. Hopefully, this will enable a concerted effort towards characterising the multi-scale, multi-physics nature of planet formation and evolution within the context of the host galaxy.

\acknowledgments

\vspace{-0.5cm}
\noindent We thank Walter Dehnen, Jaeyeon Kim, Maya Petkova, and Sebastian Trujillo-Gomez for helpful discussions. J.M.D.K.\ and M.C.\ gratefully acknowledge funding from the Deutsche Forschungsgemeinschaft (DFG, German Research Foundation) through an Emmy Noether Research Group (grant number KR4801/1-1) and the DFG Sachbeihilfe (grant number KR4801/2-1). J.M.D.K., M.C., and B.W.K.\ gratefully acknowledge funding from the European Research Council (ERC) under the European Union's Horizon 2020 research and innovation programme via the ERC Starting Grant MUSTANG (grant agreement number 714907). C.F.P.L.\ acknowledges funding from the European Research Council (ERC) under the European Union’s Horizon 2020 research and innovation program (grant agreement number 852839). This research made use of data from the European Space Agency mission \textit{Gaia} (\href{http://www.cosmos.esa.int/gaia}{http://www.cosmos.esa.int/gaia}), processed by the \textit{Gaia} Data Processing and Analysis Consortium (DPAC, \href{http://www.cosmos.esa.int/web/gaia/dpac/consortium}{http://www.cosmos.esa.int/web/gaia/dpac/consortium}). Funding for the DPAC has been provided by national institutions, in particular the institutions participating in the \textit{Gaia} Multilateral Agreement. This research has made use of the NASA Exoplanet Archive, which is operated by the California Institute of Technology, under contract with the National Aeronautics and Space Administration under the Exoplanet Exploration Program.

\software{
\package{matplotlib} \citep{hunter07},
\package{numpy} \citep{vanderwalt11},
\package{pandas} \citep{reback20},
\package{scipy} \citep{jones01},
\package{seaborn} \citep{waskom20}
}

\bibliographystyle{aasjournal}
\bibliography{mybib}

\end{document}